\documentclass[%
 reprint,
nofootinbib,
 amsmath,amssymb,
 aps,
 prd,
]{revtex4-1}

\usepackage{graphicx}
\usepackage{dcolumn}
\usepackage{bm}
\usepackage{epsfig}
\usepackage{amsmath}
\usepackage{enumerate}
\usepackage{mathrsfs}
\usepackage{natbib}
\usepackage{hyperref}
\usepackage{hhline}
\usepackage{color}

\newcommand\lsim{\mathrel{\rlap{\lower4pt\hbox{\hskip1pt$\sim$}}
        \raise1pt\hbox{$<$}}}
\newcommand\gsim{\mathrel{\rlap{\lower4pt\hbox{\hskip1pt$\sim$}}
        \raise1pt\hbox{$>$}}}

\begin{document}
\preprint{APS/123-QED}

\title{Oscillations and Random Walk of the Soliton Core in a Fuzzy Dark Matter Halo}

\author{Xinyu Li}
\email{xli@cita.utoronto.ca}
\affiliation{%
Canadian Institute for Theoretical Astrophysics, 60 St George St, Toronto, ON M5R 2M8\\
Perimeter Institute for Theoretical Physics, 31 Caroline Street North, Waterloo, Ontario, Canada, N2L 2Y5}
\author{Lam Hui}
 \email{lhui@astro.columbia.edu}
\affiliation{
Center for Theoretical Physics, Department of Physics, Columbia University, New York, NY 10027
}
\author{Tomer D. Yavetz}
\email{t.yavetz@columbia.edu}
\affiliation{Department of Astronomy, Columbia University, New York, NY 10027}%

\date{\today}

\begin{abstract}
A Fuzzy Dark Matter (FDM) halo consists of a soliton core close to the center and an NFW-like density profile in the outer region.
Previous investigations found that the soliton core exhibits temporal oscillations and random walk excursions around the halo center.
Analyzing a set of numerical simulations, we show that both phenomena can be understood as the results of wave interference---a suitable superposition of the ground (solitonic) state and excited states in a fixed potential suffices to account for the main features of these phenomena. 
Such an eigenmode analysis can shed light on the evolution of a satellite halo undergoing tidal disruption. As the outer halo is stripped away, reducing the amplitudes of the excited states, the ground state evolves adiabatically. This suggests diminished soliton oscillations and random walk excursions, an effect to consider in deducing constraints from stellar heating.
\end{abstract}

\maketitle


\section{\label{sec:level1}Introduction}

Despite the rich astronomical evidence for dark matter, its basic properties remain mysterious. An example is the mass of its constituents. Proposals ranging from $10^{-22}$ eV to solar mass can be found in the literature. In this article, we are interested in the ultra-light end of the spectrum $10^{-22} - 10^{-20}$ eV, known as fuzzy dark matter (FDM) \cite{Hu:2000ke} (see also \cite{Baldeschi:1983mq,Turner:1983he,Press:1989id,Sin:1992bg,Goodman:2000tg,Peebles:2000yy,Lesgourgues:2002hk,Amendola:2005ad,Chavanis:2011zi}). A concrete realization is an axion, or axion-like-particle, whose relic abundance is determined by the misalignment mechanism. 
It can be shown the relic abundance from the big bang matches that for dark matter if the misalignment angle is order unity, the axion decay constant is between Grand Unification and Planck scales, and the axion mass is around $10^{-22}$ eV
\cite{Arvanitaki:2009fg,Marsh:2015xka,Hui:2016ltb}.
\footnote{It is worth emphasizing the relic abundance is more sensitive to the choice of the axion decay constant $F$ than to the axion mass $m_a$, scaling as $F^2 m_a^{1/2}$. There is thus a rather large possible range for $m_a$.}
In the non-relativistic regime, the Klein-Gordon equation for the real axion field \footnote{Self-interaction for the axion can be largely ignored as far as structure formation is concerned.} can be recast as Schr\"{o}dinger equation for a complex scalar $\psi$:
\begin{eqnarray}\label{eqn:sch}
i \hbar \partial_t
  \psi=\left(-\frac{\hbar^2}{2m_a}\nabla^2+m_a V\right)\psi \, ,
\end{eqnarray}
where $m_a$ is the mass of the particle,
and the gravitational potential $V$ obeys the Poisson equation
\footnote{For a version of these equations accounting for cosmic expansion, see e.g. \cite{Hui:2016ltb}.}
\begin{eqnarray}\label{eqn:poisson}
\nabla^2 V = 4\pi G (\rho-\bar{\rho}) \, .
\end{eqnarray}
Here, $\rho$ is the mass density, related to $\psi$ by $\rho = m_a |\psi|^2$ and $\bar{\rho}$ is the mean density. In other words, $|\psi|^2$ expresses the number density of particles. We will follow custom and refer to $\psi$ as the wavefunction, though it is worth stressing that $\psi$ is a {\it classical} complex scalar field. We are interested in a regime where there are many particles per de Broglie volume, such that quantum fluctuations are negligible. 
\footnote{The de Broglie wavelength is 
$\lambda_{\rm dB} = 2\pi \hbar/(m_a v) = 0.5 {\,\rm kpc} (m_a / 10^{-22} {\,\rm eV})^{-1} (v / 250 {\,\rm km/s})^{-1}$. It can be shown the number of particles in a de Broglie volume is roughly $(30 {\,\rm eV}/m_a)^4$ for a dark matter density around the solar neighborhood value.
}
The situation is similar to the one in electromagnetism: a state with many photons is often well described by the classical electric and magnetic fields.
\footnote{One might wonder how to think about the factors of $\hbar$ in our {\it classical} wave setting.
Imagine dividing the Schr\"odinger equation by $m_a$. One sees that $\hbar$ always comes in the combination $\hbar/m_a$. This quantity has the dimension of ${\,\rm length}^2 / {\,\rm time}$. At a  pragmatic level, in terms of solving the equation, one could think of $\hbar/m_a$ as just some given quantity with this dimension. Of course, ultimately this quantity is connected with the mass of the particle in question, and the relation involves $\hbar$.
}

The Schr\"odinger-Poisson system can be interpreted as describing a fluid, where the fluid density $\rho$ and velocity $\vec v$ are defined by \citep{Madelung1927}:
\begin{equation}
	\psi\equiv\sqrt{\frac{\rho}{m_a}}e^{i\theta} \quad, \quad
{\vec v}\equiv \frac{\hbar}{m_a}\vec \nabla \theta \, .
\end{equation}
The Schr\"odinger equation can be rewritten as
\begin{eqnarray}\label{eqn:fluid}
	\partial_t \rho + \vec \nabla\cdot(\rho\vec {v})&=&0 \, ,\\
	\partial_t \vec {v} + (\vec {v}\cdot\vec \nabla)\vec {v} &=& -\vec \nabla V + \frac{\hbar^2}{2m_a^2}\vec \nabla {\nabla^2 \sqrt{\rho} \over \sqrt{\rho}}\, .
\end{eqnarray}
The first expresses mass conservation, while the second is the Euler equation. The last term of the Euler equation is often referred to as the quantum pressure term, though it arises from a non-trivial stress tensor rather than mere pressure. It is worth stressing that while the fluid formulation is useful for understanding many aspects of the Schr\"odinger-Poisson system, it fails at locations where $\rho$ vanishes. These are sites of vortices which occur at a rate of roughly one vortex ring per de Broglie volume \cite{Hui:2020hbq}.

There is a substantial literature devoted to the study of structure formation in FDM using numerical simulations, starting from the work of Schive, Chiueh, and Broadhurst \cite{Schive:2014dra,Schive:2014hza,Mocz:2015sda,Mocz:2017wlg,Zhang:2016uiy,Schwabe:2016rze,Li:2018kyk,Nori:2018hud,Veltmaat:2016rxo,Veltmaat:2018dfz,Schwabe:2020eac}. It is found that an FDM halo has an outer NFW-like \cite{Navarro:1996gj} density profile, and a distinctive central core with the profile 
\cite{Schive:2014dra,Schive:2014hza}:
\begin{equation}\label{eqn:sol_dens}
    \rho_c (r) = \frac{0.019(r_c/\,\mathrm{kpc})^{-4} }{[1+0.091(r/r_c)^2]^8}\left(\frac{m_a}{10^{-22}\,\mathrm{eV}}\right)^{-2}\,M_\odot\mathrm{pc}^{-3}.
\end{equation}
The core resembles a soliton or boson star---a gravitationally bound object supported by quantum pressure. Its properties can be deduced by balancing gravity against quantum pressure (the two terms on the right hand side of the Euler equation), giving roughly
\begin{equation}
    \frac{G M_c}{r_c} \sim  \frac{\hbar^2}{m_a^2 r_c^2}\; \quad \mathrm{or} \quad \; r_c \sim \frac{\hbar^2}{m_a^2 G M_c},
\end{equation}
where $r_c$ and $M_c$ are the radius and mass of the soliton. This is why the density $\rho_c$ scales as $r_c^{-4} m_a^{-2}$ or $M_c^4 m_a^6$.

A soliton is strictly speaking a stationary state, that is, the corresponding wavefunction has a time dependence that resides entirely in its phase --- the associated density should be time independent. It is thus a very interesting finding by Veltmaat, Niemeyer, and Schwabe \cite{Veltmaat:2018dfz} that the central density of an isolated FDM halo oscillates in time, with an order unity amplitude.
The central core of an FDM halo is hence not a strict soliton, but a perturbed one. This phenomenon was used to constrain FDM
based on the existence of a stellar cluster close to the center of Eridanus II \cite{Marsh:2018zyw}: oscillations of the soliton (or more precisely, the perturbed soliton) have the potential to completely disrupt the stellar cluster. 
It was subsequently pointed out by \cite{Schive:2019rrw} that not only does the soliton oscillate, it also random walks around the central region of the halo. It was also shown that if the halo is a satellite of some larger parent halo, as is in the case of Eridanus II, the random walk excursions are diminished in amplitude after accounting for tidal stripping, alleviating the constraint from stellar heating. 

In this article, we offer an interpretation of the soliton oscillation and random walk phenomena based on wave interference. The idea is a simple one: think of the halo wavefunction as composed of a superposition of energy eigenstates, schematically of the form 
$\psi = \sum_p a_p \psi_p e^{-i E_p t /\hbar}$ where $\psi_p$'s represent the energy eigenstates labeled by $p$ with amplitudes $a_p$'s.
When the wavefunction is ``squared'' to obtain the density, time dependence arises from cross terms that involve different eigenmodes, in other words interference terms.
In particular, interference between the ground state and excited states explains the oscillation and random walk phenomena, as we will see. 
It is not a priori obvious this is a fruitful way of thinking. The issue is that the gravitational potential itself fluctuates in time, and thus the time dependence is not completely captured by the phase factor of $e^{-i E_p t / \hbar}$ for each state (i.e. effectively, $a_p \psi_p$ fluctuates with time). At a detailed level, this is undoubtedly true, but we will see that modeling the FDM halo as a superposition of energy eigenstates in a {\it fixed} potential is a reasonable first approximation. 
We demonstrate this by performing numerical simulations of halos from gravitational collapse, and carrying out an eigenmode analysis of them. The goal is a unified, wave-interference explanation of both the oscillation and random walk phenomena of the central soliton in an FDM halo. For an earlier application of the eigenmode technique to construct FDM halos, see \cite{Lin:2018whl}. See also \cite{Padmanabhan:2020} for an independent analysis that overlaps with ours.

\section{Numerical Simulations and Eigenmode Analysis}
\label{sec:sim}
Our numerical simulations utilize the \texttt{SPoS} code presented and tested in \citep{Li:2018kyk}. 
The FDM mass is chosen to be $m_a=10^{-22}$ eV. The box size is $25$ kpc on a $256^3$ grid, giving a resolution of about $0.1$ kpc.
These are not cosmological simulations, but rather simulations of the formation of isolated halos from gravitational collapse---we study a case of a spherically symmetric collapse, and a case involving the collision of seed solitons. 
In both cases, the final halo mass is $1-2 \times 10^9 M_\odot$. 
We have checked that the corresponding de Broglie wavelength is resolved. 
Periodic boundary conditions are imposed, though we have verified simulations with absorbing boundary conditions yield similar results.

It is worth noting that the simulations can be rescaled to describe halos with different masses and sizes. The Schr\"{o}dinger-Poisson system is invariant under the following Lifshitz-type transformation
\begin{equation}\label{eqn:scaling}
    \{t, x, V, \psi, \rho \} \rightarrow \{ \lambda t, \lambda^{1/2}x, \lambda^{-1} V, \lambda^{-1}\psi, \lambda^{-2} \rho\} \, ,
\end{equation}
where $\lambda$ is the scaling parameter. The ratio of de Broglie wavelength to halo size is invariant under this transformation, as is the product of soliton size and soliton mass. The FDM mass $m_a$ is held fixed in this transformation. 
One could also contemplate rescaling $m_a$:
\begin{equation}
    \{m_a, t, x, V, \psi,\rho\} \rightarrow \{\alpha m_a, \alpha t, x, \alpha^{-2} V, \alpha^{-3/2} \psi, \alpha^{-2}\rho\} \, ,
\end{equation}
with $\alpha$ as the scaling parameter. 

\subsection{Collapse of a Spherically Symmetric Halo}
Our first numerical experiment starts with an initial spherical top-hat density profile:
\begin{equation}
    \rho = 0.137 \, \Theta(r_0-r)\,M_\odot\mathrm{pc}^{-3} \, ,
\end{equation}
where $r_0 = 1.25$~kpc and $\Theta$ is the Heaviside step function.
The initial wave function is taken to be real $\psi=\sqrt{\rho/m_a}$ i.e. no initial velocity. 

\begin{figure}[htpb]
\includegraphics[width=0.45\textwidth]{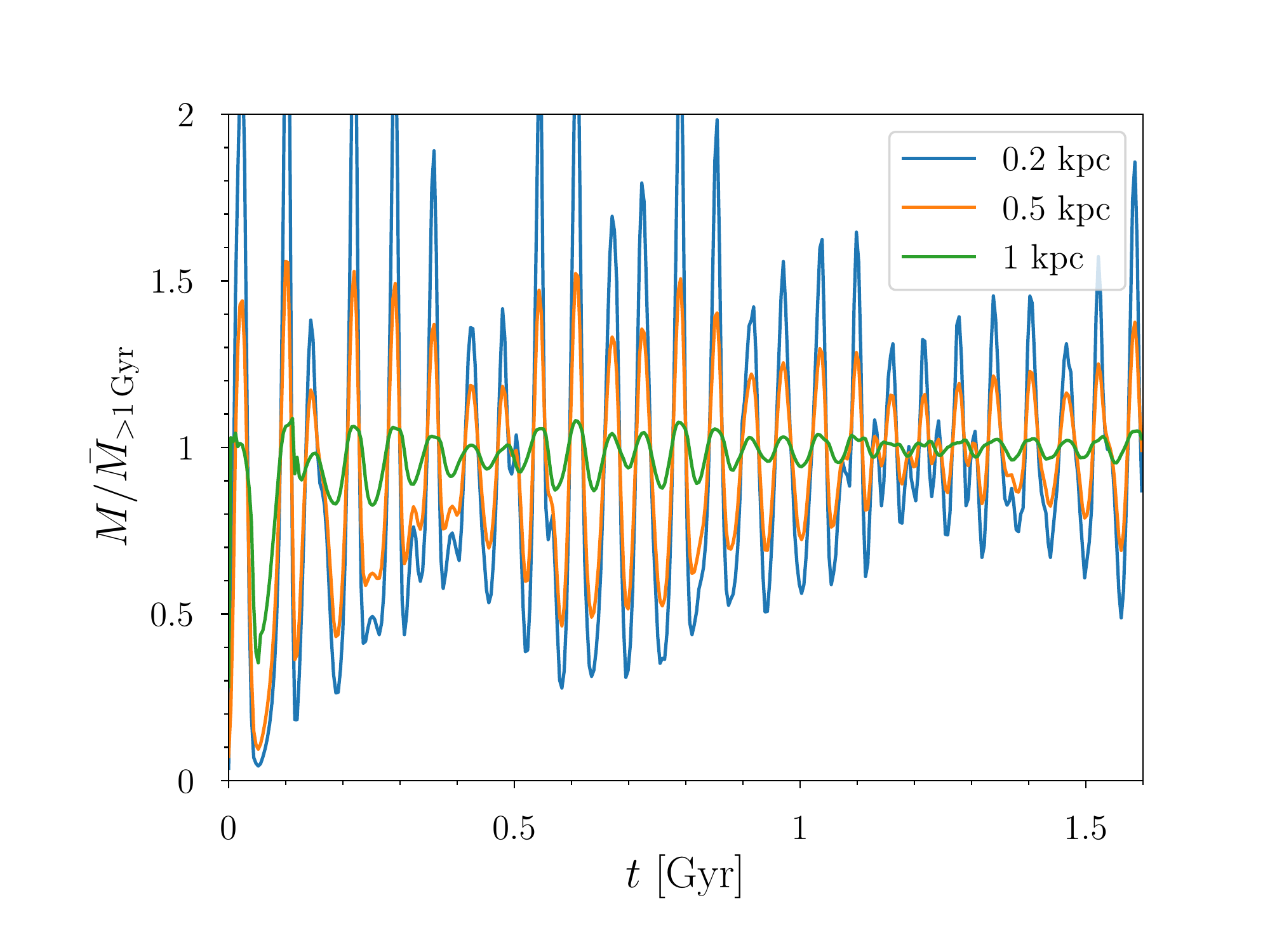} 
\caption{Evolution of total mass enclosed inside different radii from the halo center in the simulation of a spherically symmetric collapse.
}
\label{fig:osc_tophat}
\end{figure}

Gravitational collapse leads to the formation of a halo with a central core. 
Figure~\ref{fig:osc_tophat} shows the evolution of total mass enclosed inside different radii ($0.2$~kpc, $0.5$~kpc and $1$~kpc) from the halo center. 
The $y$-axis shows the enclosed mass normalized by the average value between $1-1.6$~Gyr for each radius.
Large oscillations commence early on, settling down to order unity oscillations (within $0.2$ kpc) after about $1$ Gyr. 
The oscillation period is about $0.05$ Gyr.

\begin{figure}[htpb]
\includegraphics[width=0.45\textwidth]{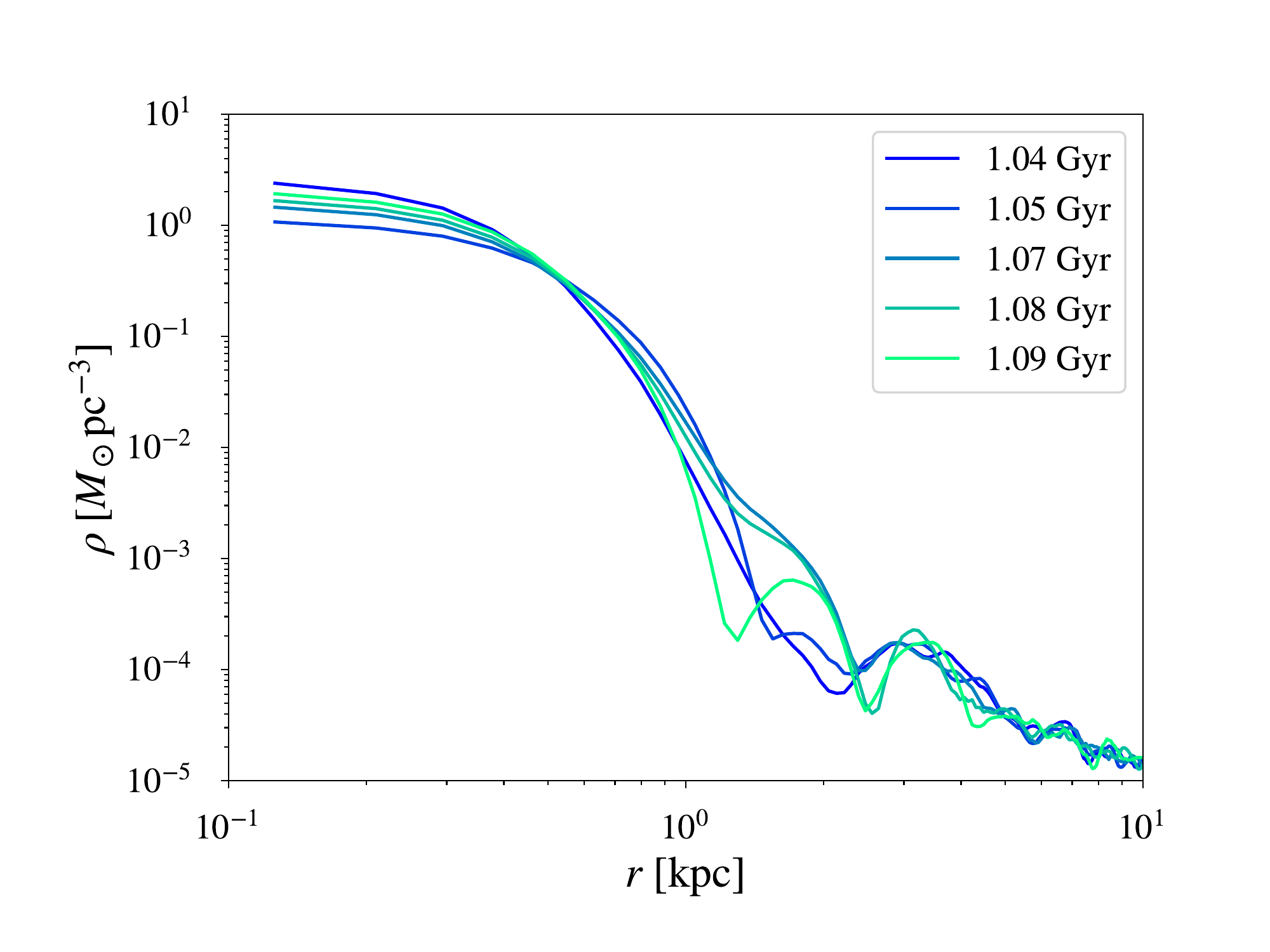} 
\includegraphics[width=0.45\textwidth]{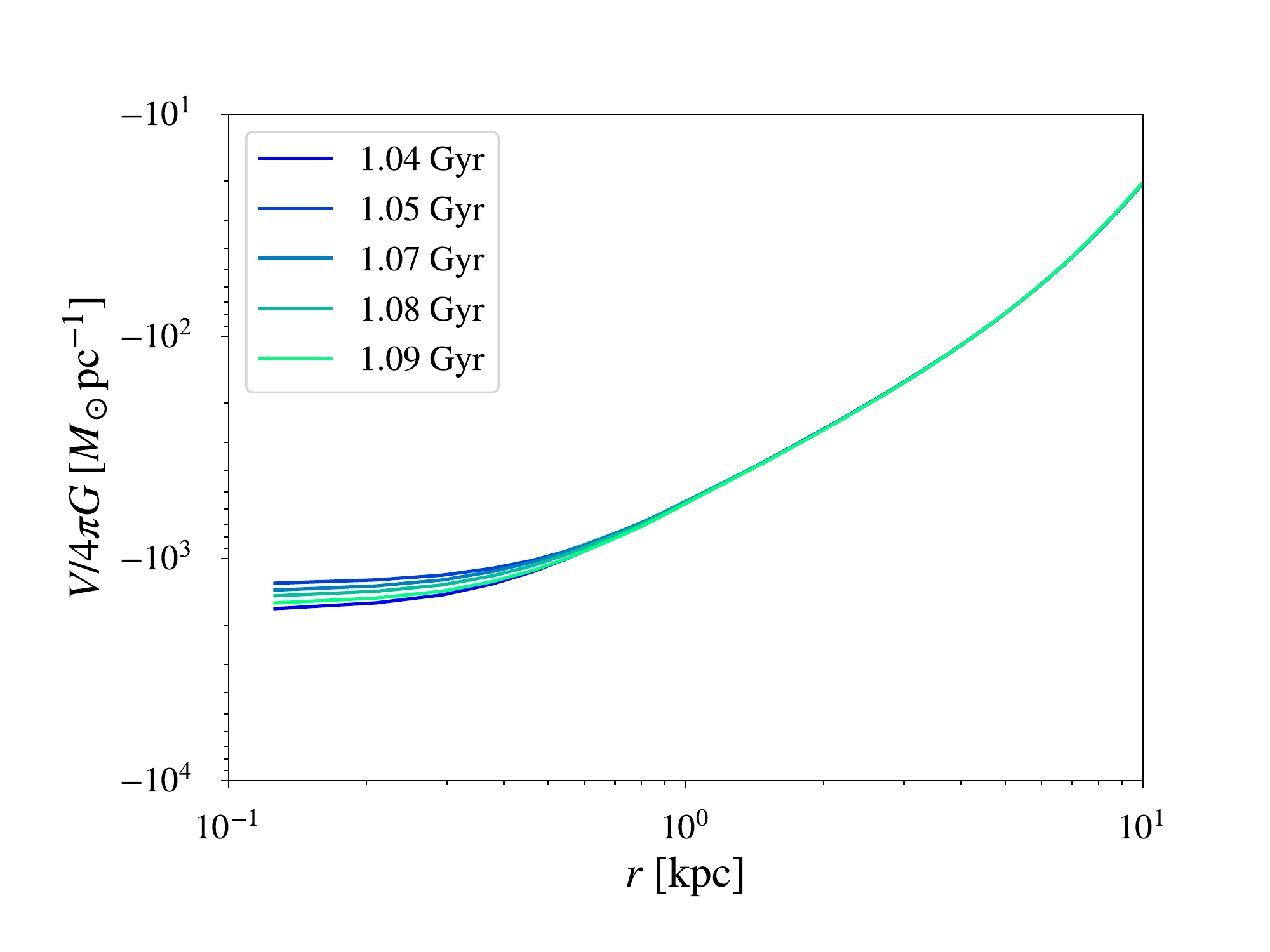} 
\caption{Snapshots of the halo density (upper panel) and gravitational potential (lower panel) profiles over an oscillation period. This is for the halo that forms from a spherically symmetric collapse.
}
\label{fig:profile_tophat}
\end{figure}

Figure~\ref{fig:profile_tophat} offers a more detailed view of what is going on. The two panels show respectively the density and gravitational potential profiles at several different instants that span over an oscillation period.
The central density profile reveals a soliton-like object. In this case of spherical symmetry, the halo center and the soliton center coincide. The soliton (or more properly, the {\it perturbed} soliton) exhibits oscillations like a pulsating sphere. 
The gravitational potential displays less variation with time compared to the density. This is to be expected: the potential is sourced by the density over many locations and is thus smoother. 
The largest variations in potential occur at small radii, and even there, the oscillation amplitude is less than 10 percent.

\begin{figure}[htpb]
\includegraphics[width=0.45\textwidth]{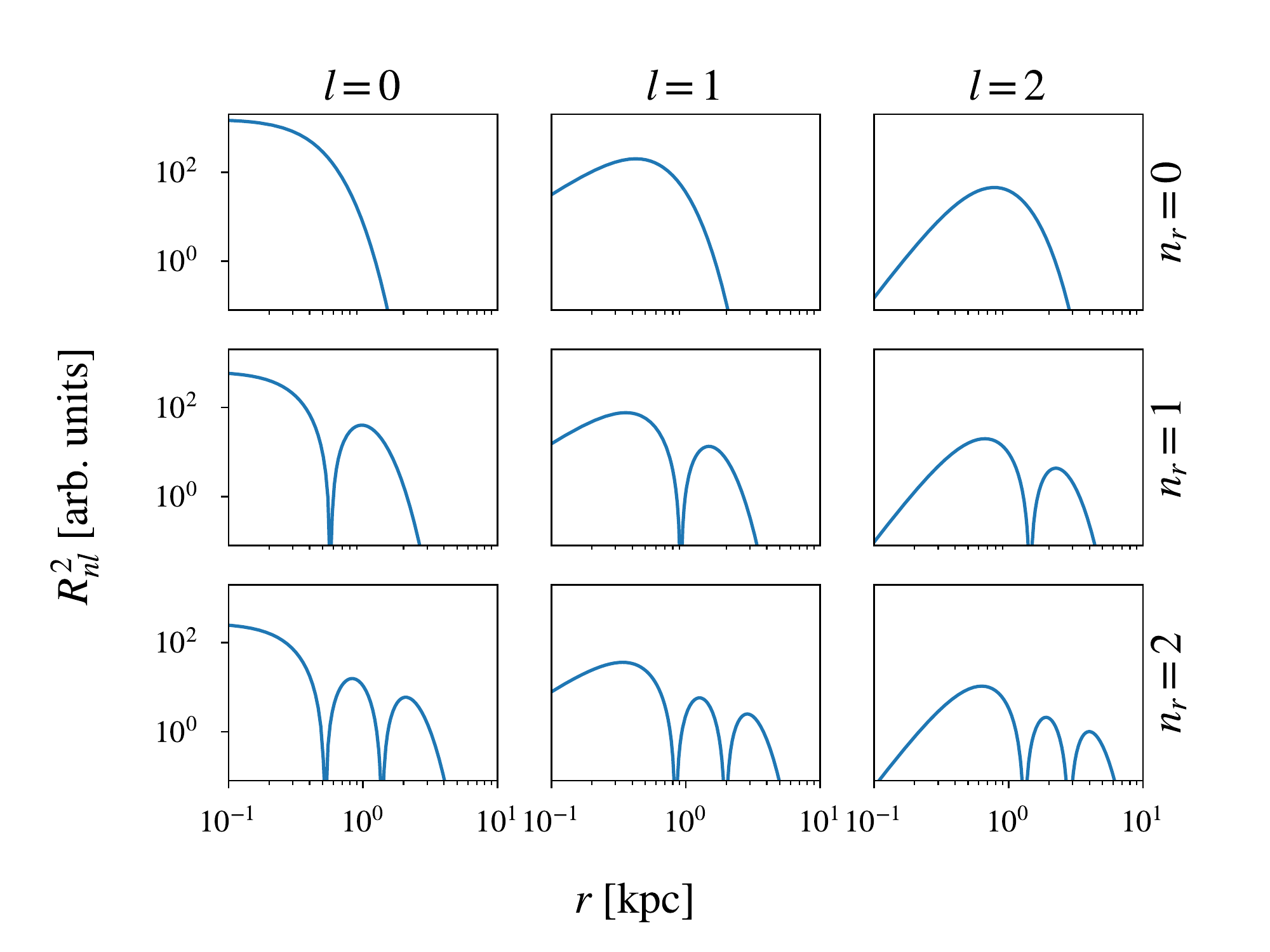} 
\caption{9 radial eigenfunctions $R_{nl}$ with lowest energies for the averaged potential. $n_r$ is the radial quantum number which counts the number of nodes in $R_{nl}$ with $n=n_r+l+1$.
}
\label{fig:eigen}
\end{figure}

This motivates us to perform an eigenmode decomposition of the halo wavefunction, where the eigenmodes are for a {\it fixed} gravitational potential --- we use the average potential between $1 - 1.6$ Gyr. The eigenfunctions are labeled by $n, l, m$:
\begin{equation}
    F_{nlm}(r,\theta,\phi)=R_{nl}(r)Y_l^{m} (\theta,\phi) \, ,
\end{equation}
where $Y_l^{m}$ denotes the spherical harmonics, and $R_{nl}$ is the radial eigenfunction. The use of spherical harmonics is not strictly necessary for a spherically symmetric situation, but will be useful for more general cases. Each eigenfunction satisfies:
\begin{eqnarray}
    && E_{nl} R_{nl} = -{\hbar^2 \over 2m_a r^2} {d \over dr} \left(r^2 {dR_{nl} \over dr}\right) \nonumber \\
    && \quad \quad \quad \quad +\left[ {\hbar^2 \over 2m_a r^2} {l(l+1)} + m_a V \right] R_{nl} \, ,
\end{eqnarray}
where $E_{nl}$ is the energy eigenvalue, and $V$ is the average (fixed) potential.
The eigenfunctions are properly normalized in the sense that
\begin{equation}
    \int r^2 {\,\rm sin}\theta \,d\theta \, d\phi \,  dr \, |F_{nlm}|^2 = 1 \, ,
\end{equation}
where we integrate over the whole computational box. 
Figure ~\ref{fig:eigen} shows 9 radial eigenfunctions with the smallest $l$ and $n_r$ which we numerically obtain.
Here $n_r$ is the radial quantum number which counts the number of nodes in $R_{nl}$; it is related to the principle quantum number $n$ through $n=n_r+l+1$.

The halo wavefunction $\psi$ is decomposed as a superposition of the eigenmodes:
\begin{equation}
    \psi(r,\theta,\phi, t) = \sum_{n,l,m} A_{nlm} F_{nlm} (r,\theta,\phi) e^{-iE_{nl} t/\hbar} \, .
\label{psidecompose}
\end{equation}

\begin{figure}[htpb]
\includegraphics[width=0.45\textwidth]{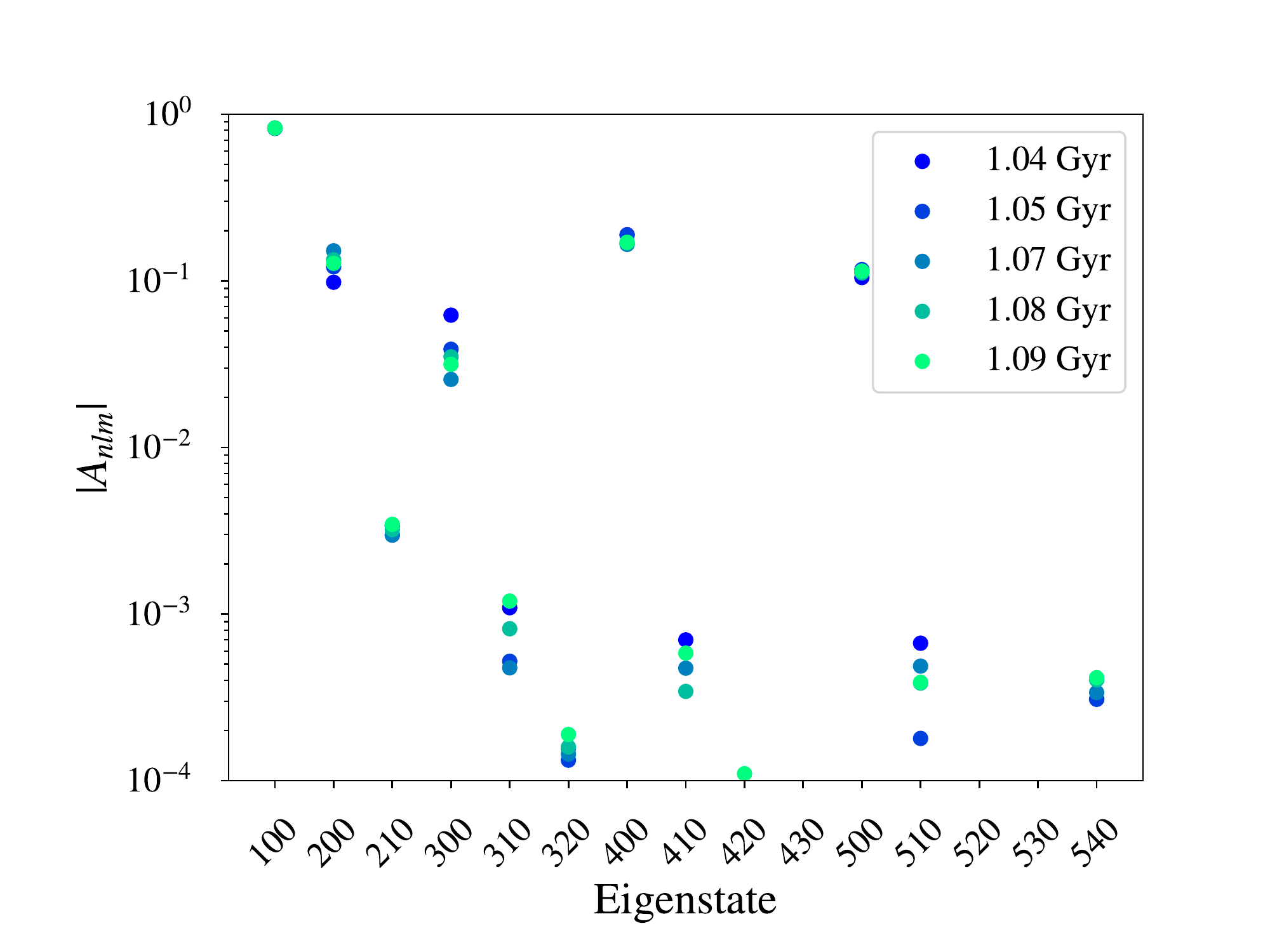} 
\caption{The superposition coefficients $A_{nlm}$ at different moments in time during one period of oscillation, for the halo that forms from spherical collapse.
The eigenstates are labeled by $nlm$.}
\label{fig:modes_tophat}
\end{figure}

Figure~\ref{fig:modes_tophat} shows the superposition coefficients $A_{nlm}$ for all eigenstates with $n\leq 5$. Here, and in the figures in the rest of the paper, we show $A_{nlm}$ divided by $(\int d^3 x \, |\psi|^2)^{1/2}$. 
The precise superposition depends on the time at which $\psi$ is decomposed, and the figure displays several different snapshots over an oscillation period.
The wavefunction $\psi$ has the largest projection onto the ground state $(nlm) = (100)$. 
The coefficient $A_{100}$ is also the most stable over time. The next dominant modes are the excited states $(nlm) = (200)$ and $(400)$. 
Modes with non-zero $l$'s have small amplitudes by virtue of the spherically symmetric initial conditions; their amplitudes do not exactly vanish because of numerical noise and the cubic box boundary condition. 
(Modes with non-zero $m$'s have even smaller coefficients, which are not shown.)
We have alternatively carried out the decomposition using the eigenfunctions corresponding to the gravitational potential at each snapshot---the results are similar to what is shown in Figure \ref{fig:modes_tophat}.

\begin{figure}[htpb]
\includegraphics[width=0.45\textwidth]{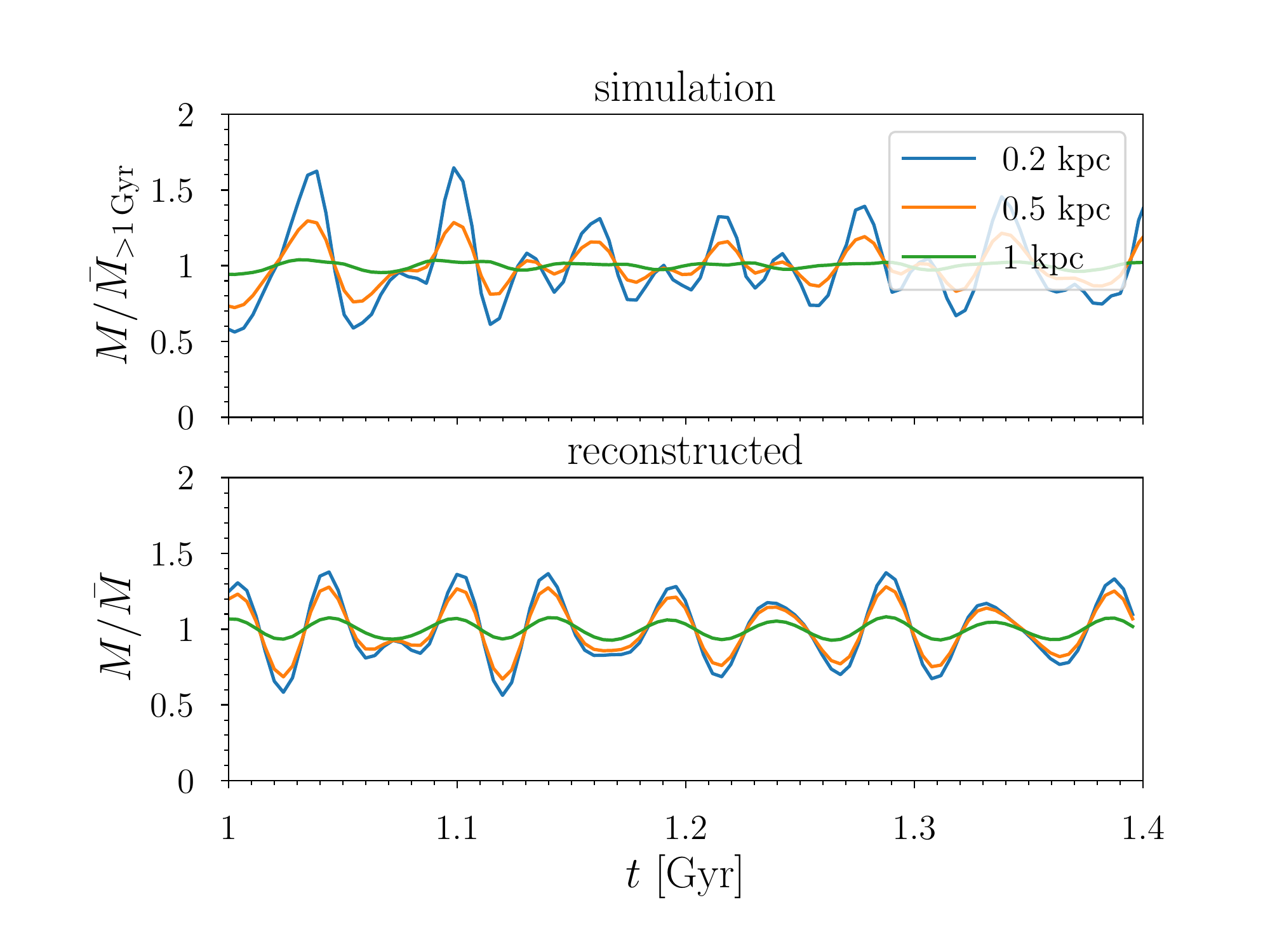}
\includegraphics[width=0.45\textwidth]{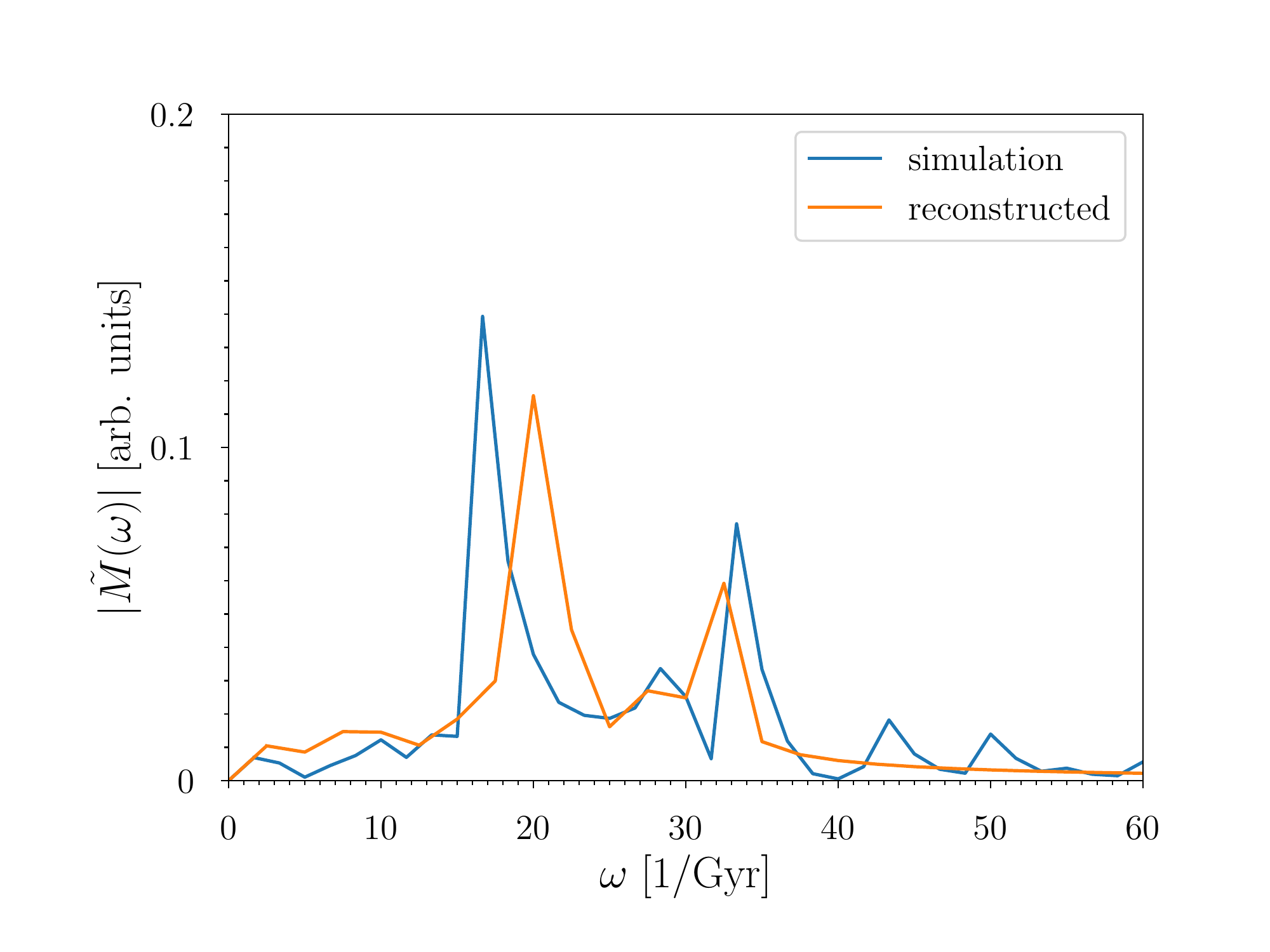} 
\caption{Comparison of soliton oscillation measured from the simulation against that implied by the reconstructed wavefunction $\tilde{\psi}$ (Equation \ref{psidecomposeR}).
Top figure: the mass inside several different radii from the halo/soliton center.
The upper panel is from the simulation, and the lower panel is from the reconstructed model wavefunction (which includes only $l=0$ modes).
Bottom figure: Fourier coefficients $|\tilde{M}(\omega)|$ of the mass oscillation as a function of frequency $\omega$.
}
\label{fig:reconstruct_tophat}
\end{figure}

Encouraged by the relatively small temporal variation of the $A_{nlm}$'s, we reconstruct a model wavefunction $\tilde\psi$, which is none other than Equation \ref{psidecompose} except that the $A_{nlm}$'s are fixed (to be the values measured at $1.09$ Gyr):
\begin{equation}
        \tilde \psi(r,\theta,\phi, t) = \sum_{n,l,m} A_{nlm}^{\rm fix} F_{nlm} (r,\theta,\phi) e^{-iE_{nl} t/\hbar} \, ,
\label{psidecomposeR}
\end{equation}
and the eigenmodes $F_{nlm}$ are as before, computed based on the average gravitational potential between $1 - 1.6$ Gyr.
As such, the time dependence of $\tilde\psi$ arises completely from the time dependent phase factor $e^{-i E_{nl} t/\hbar}$ for each eigenmode.
Figure~\ref{fig:reconstruct_tophat} shows the comparison of soliton oscillation between the simulation and our reconstructed wave function $\tilde{\psi}$.
The upper panel in the top figure shows the mass enclosed within different radii as a function of time from the simulated halo (spanning $1 - 1.4$ Gyr as shown in Figure \ref{fig:osc_tophat}).
The lower panel in the top figure shows what the reconstructed wave function $\tilde\psi$ implies about these interior mass fluctuations. 
The two panels are broadly similar, suggesting that interference of a {\it fixed} superposition of eigenmodes, each with its $e^{-i E_{nl} t/\hbar}$ phase, is sufficient to approximately account for the observed time-variability. 
In other words, the (number) density implied by the reconstructed wave function is:
\begin{eqnarray}
    && |\tilde\psi|^2 = \sum_{n,l,m} \sum_{n',l',m'}
    A_{n'l'm'}^{\rm fix \, *} A_{nlm}^{\rm fix} \nonumber \\
    && \quad \quad \quad \quad F_{n'l'm'}^* F_{nlm} e^{-i(E_{nl} - E_{n'l'})t/\hbar}\, ,
\end{eqnarray}
where the time dependence of $|\tilde\psi|^2$ arises entirely from the phase factors associated with the cross terms $E_{nl} \ne E_{n'l'}$. 

To further quantify the agreement between simulation and reconstruction, we Fourier transform the curves of mass fluctuation (interior to $0.2$ kpc) as a function of time, and display the Fourier coefficients in the bottom figure (the $\omega = 0$ mode is removed by subtracting out the mean).
The reconstruction reproduces reasonably well the two prominent peaks observed in the fluctuation power spectrum of the simulated halo. 
The peak frequency around $20$ Gyr$^{-1}$ can be identified with $\Delta E/\hbar$, with $\Delta E$ being the energy difference between the eigenstates $(200)$ and $(100)$. The secondary peak frequency of about $33$ Gyr$^{-1}$ is associated with the energy difference between the eigenstates $(400)$ and $(100)$.
The $(300)$ eigenmode has a much smaller amplitude and can be ignored; see Figure \ref{fig:modes_tophat}.

\subsection{Collision of Solitons}

In our second numerical experiment, we study an FDM halo formed from asymmetric initial conditions. 
The initial configuration consists of 10 identical 
$r_c = 1.2$ kpc solitons (i.e., real wavefunctions, each with the density profile given by Equation~\ref{eqn:sol_dens}), placed
randomly within the computational box. Gravity causes them to fall towards each other, collide, and eventually merge. 

\begin{figure}[htpb]
\includegraphics[width=0.45\textwidth]{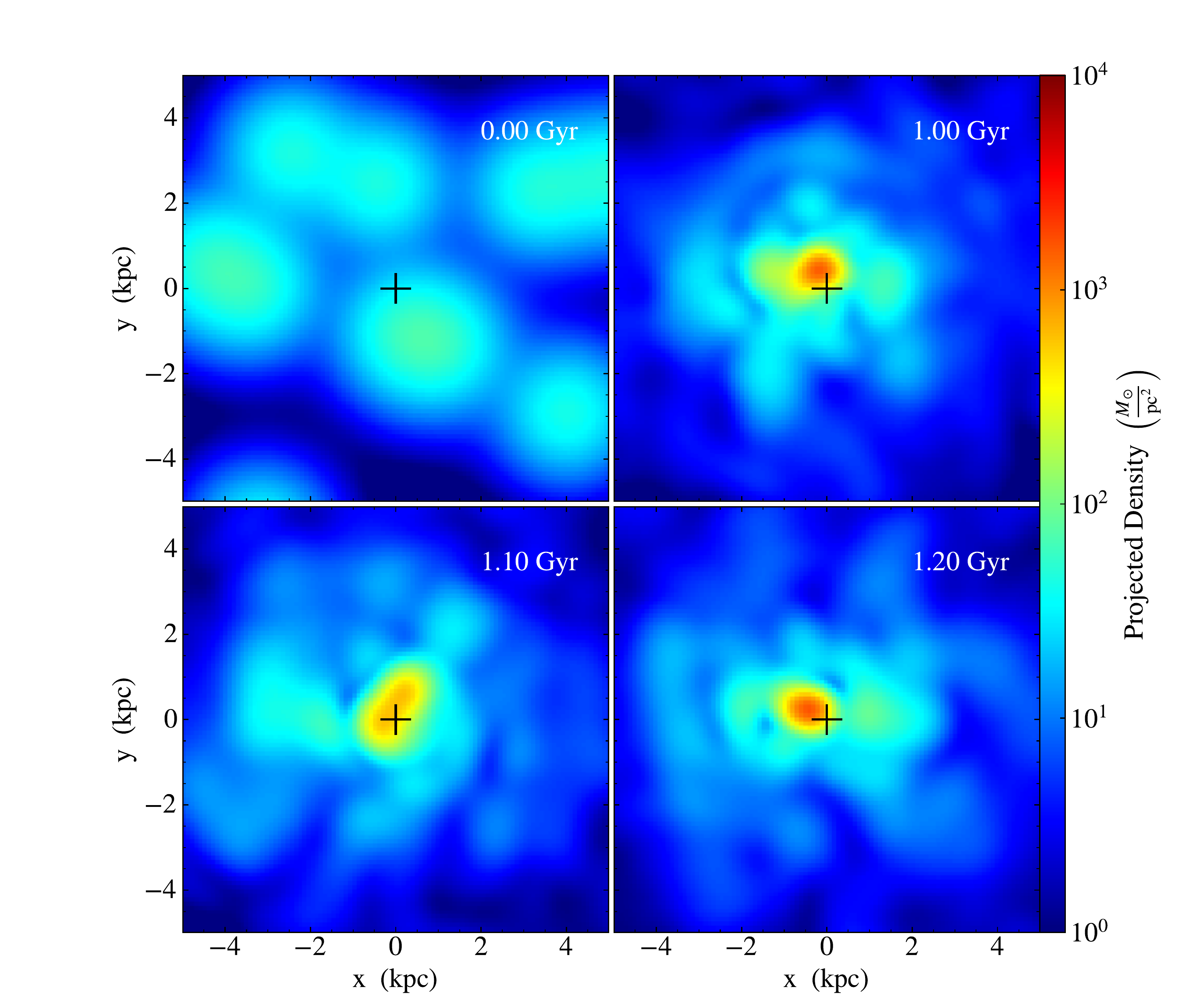} 
\caption{Snapshots of the projected density in the soliton collision simulation. The black cross in each panel denotes the center of mass.
}
\label{fig:project_col}
\end{figure}

\begin{figure}[htpb]
\includegraphics[width=0.45\textwidth]{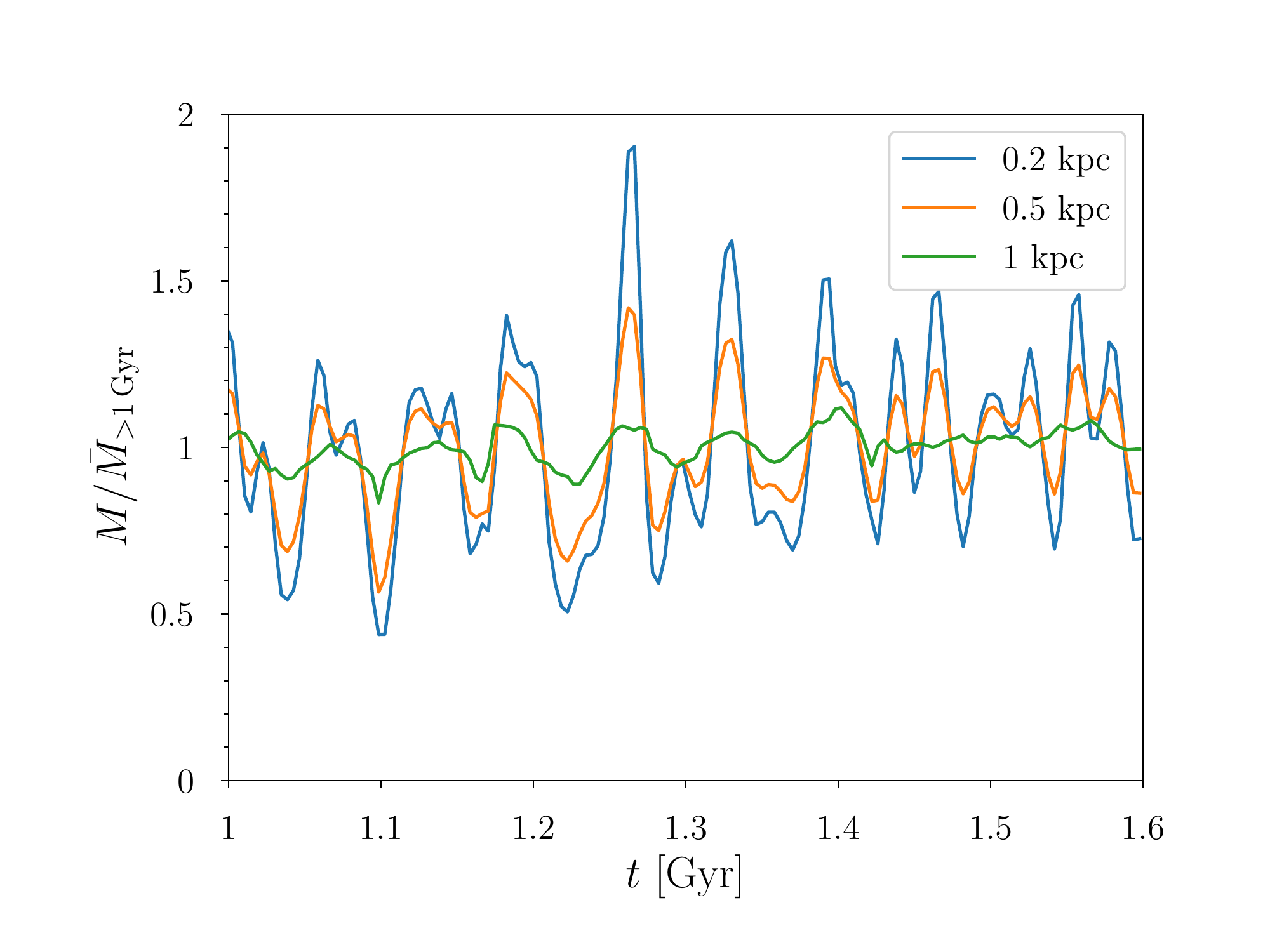} 
\caption{Evolution of mass enclosed inside different radii from the soliton center, for $t=1-1.6$~Gyr in the soliton collision simulation. Here, the mass is normalized by
the average value in the same period.
}
\label{fig:osc_col}
\end{figure}

Figure~\ref{fig:project_col} shows the zoomed-in projected density at 4 different moments of the simulation.
The merger product stabilizes after about $1$~Gyr. 
It has its own solitonic core --- the final soliton is more compact than the initial seed solitons, respecting the reciprocal relation between radius and mass (Section \ref{sec:level1}). 
The black crosses show the position of the halo center of mass; notice how the soliton center is close by, but random walks around it, consistent with the findings of \cite{Schive:2019rrw}.
(The random walk phenomenon is, by construction, absent from the spherically symmetric simulation.)
Figure~\ref{fig:osc_col} shows the variation of mass enclosed within a few radii from the soliton center.
The solitonic core reaches a more or less steady state by $1$ Gyr, with order unity oscillations. 
Just as in the spherically symmetric case,
the oscillation amplitude is smaller for larger radii.

\begin{figure}[htpb]
\includegraphics[width=0.45\textwidth]{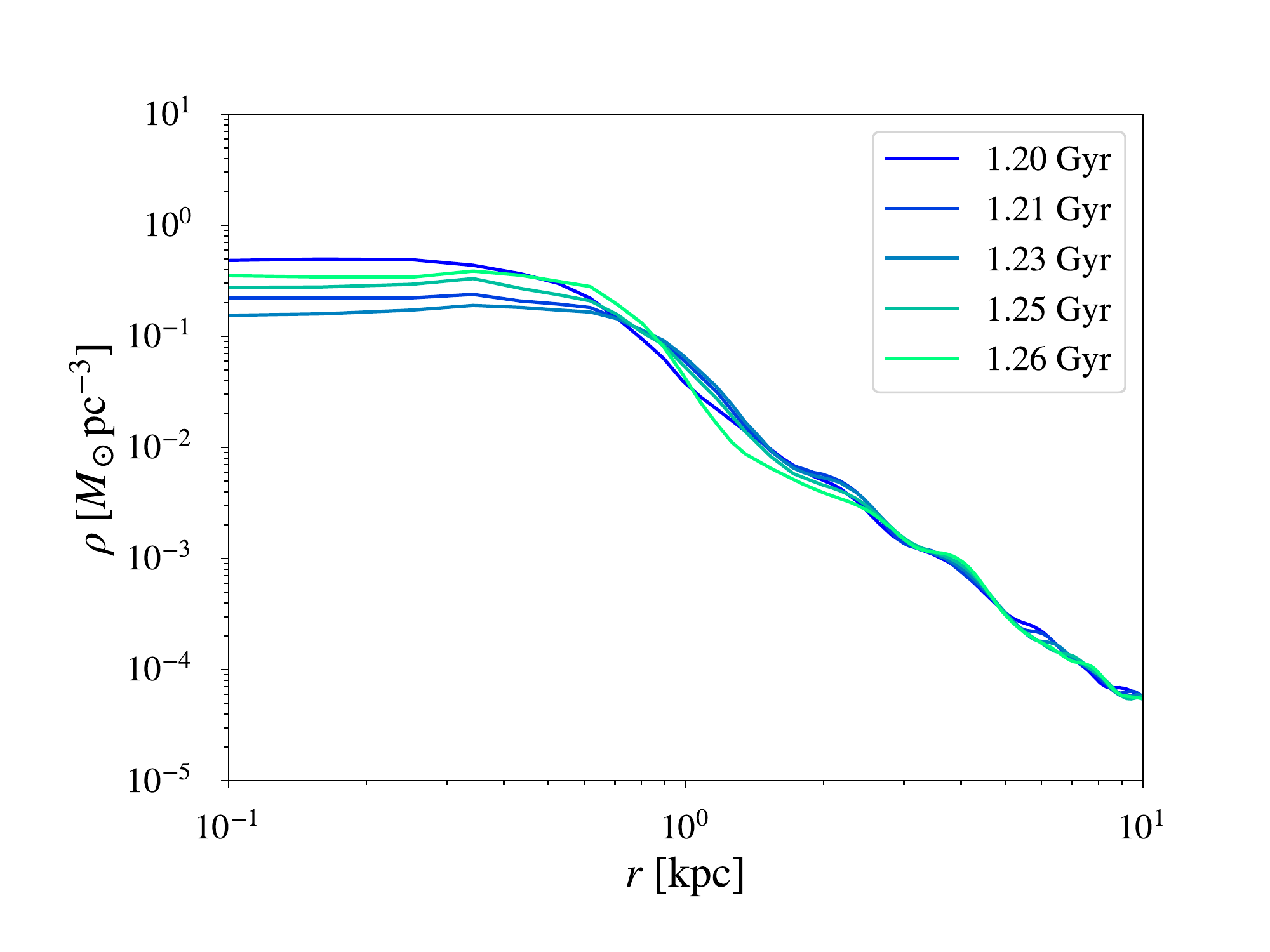} 
\includegraphics[width=0.45\textwidth]{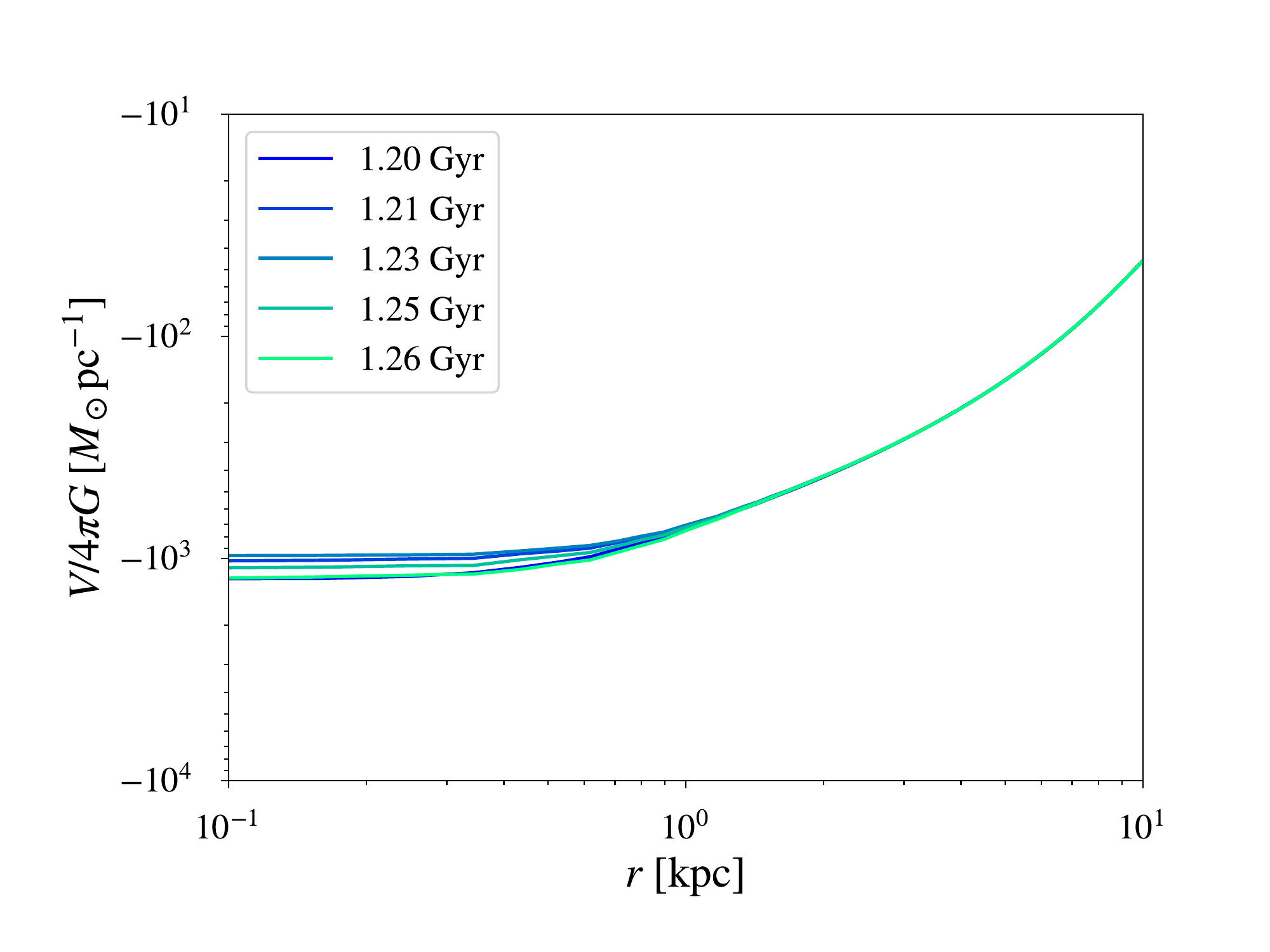} 
\caption{The density (upper panel) and gravitational potential (lower panel) profiles at several moments spanning an oscillation period, for the halo that forms from soliton collisions. Here, the profiles are spherically averaged, with $r=0$ being the halo center of mass.
}
\label{fig:profile_random}
\end{figure}

\begin{figure}[htpb]
\includegraphics[width=0.45\textwidth]{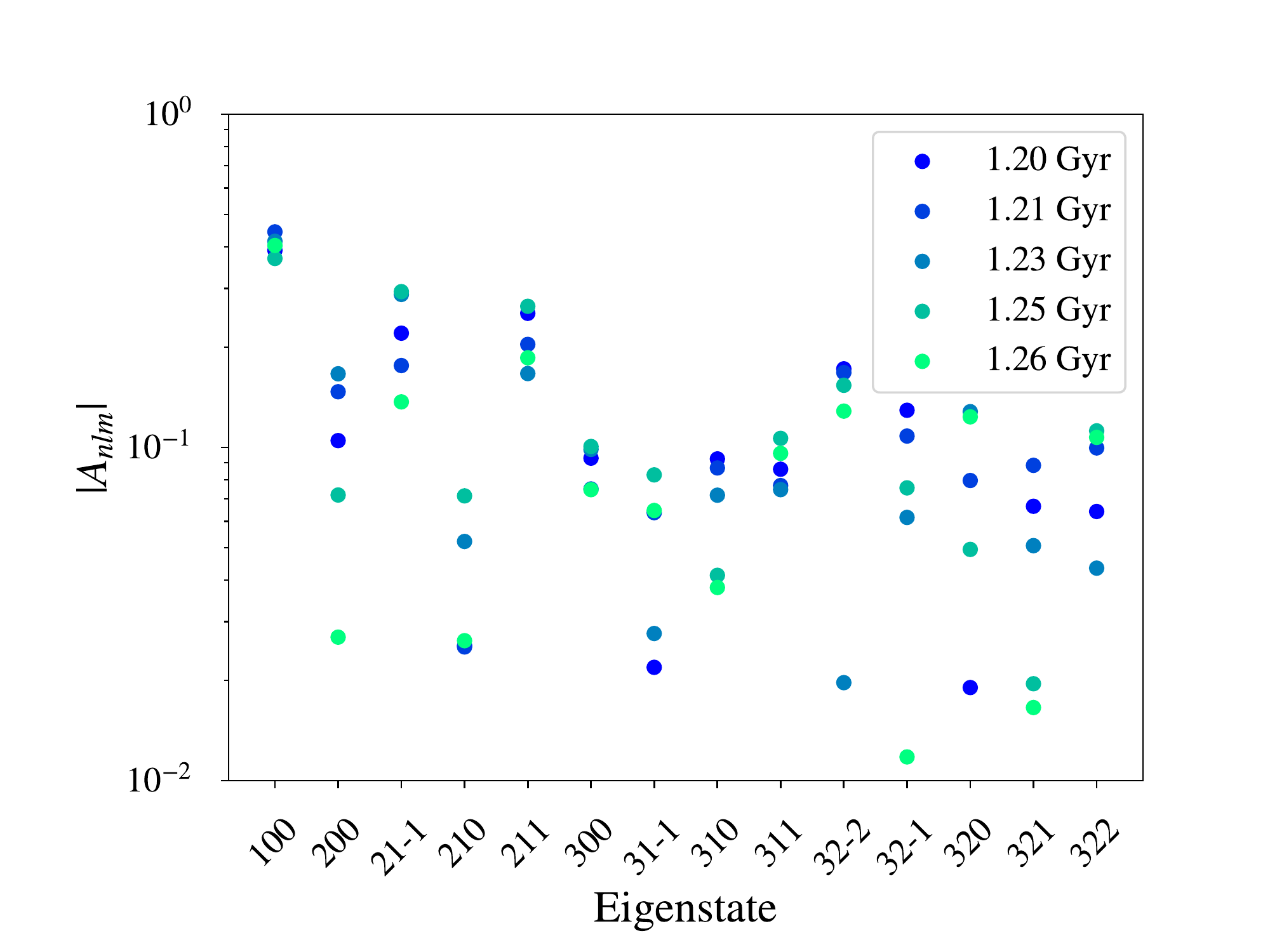} 
\caption{Coefficients of eigenstate decomposition of $\psi$ during one period of oscillation in the soliton collision simulation.
}
\label{fig:modes_random}
\end{figure}

Figure~\ref{fig:profile_random} shows the density and gravitational potential profiles of the halo where $r=0$ is the halo center of mass.
Note that these are spherically averaged profiles.
As before, both the density and the potential fluctuate with time, but the latter fluctuates less---and the fluctuations are confined to the central region. 
Encouraged by this, we compute the eigenstates using the average halo potential for $t>1$~Gyr, and perform an eigenstate decomposition of $\psi$, just as in Equation (\ref{psidecompose}). 
Figure~\ref{fig:modes_random} shows the coefficients $A_{nlm}$ for $n\leq 3$, from decomposing $\psi$ at different moments in time.
Similar to the spherically symmetric case, the ground state $(100)$ has the largest amplitude $A_{100}$, which also has the least variability.
Overall, compared to Figure \ref{fig:modes_tophat}, there are more temporal variations.
The modes with non-vanishing $l$ and $m$ are also more prominent. 

\begin{figure}[htpb]
\includegraphics[width=0.45\textwidth]{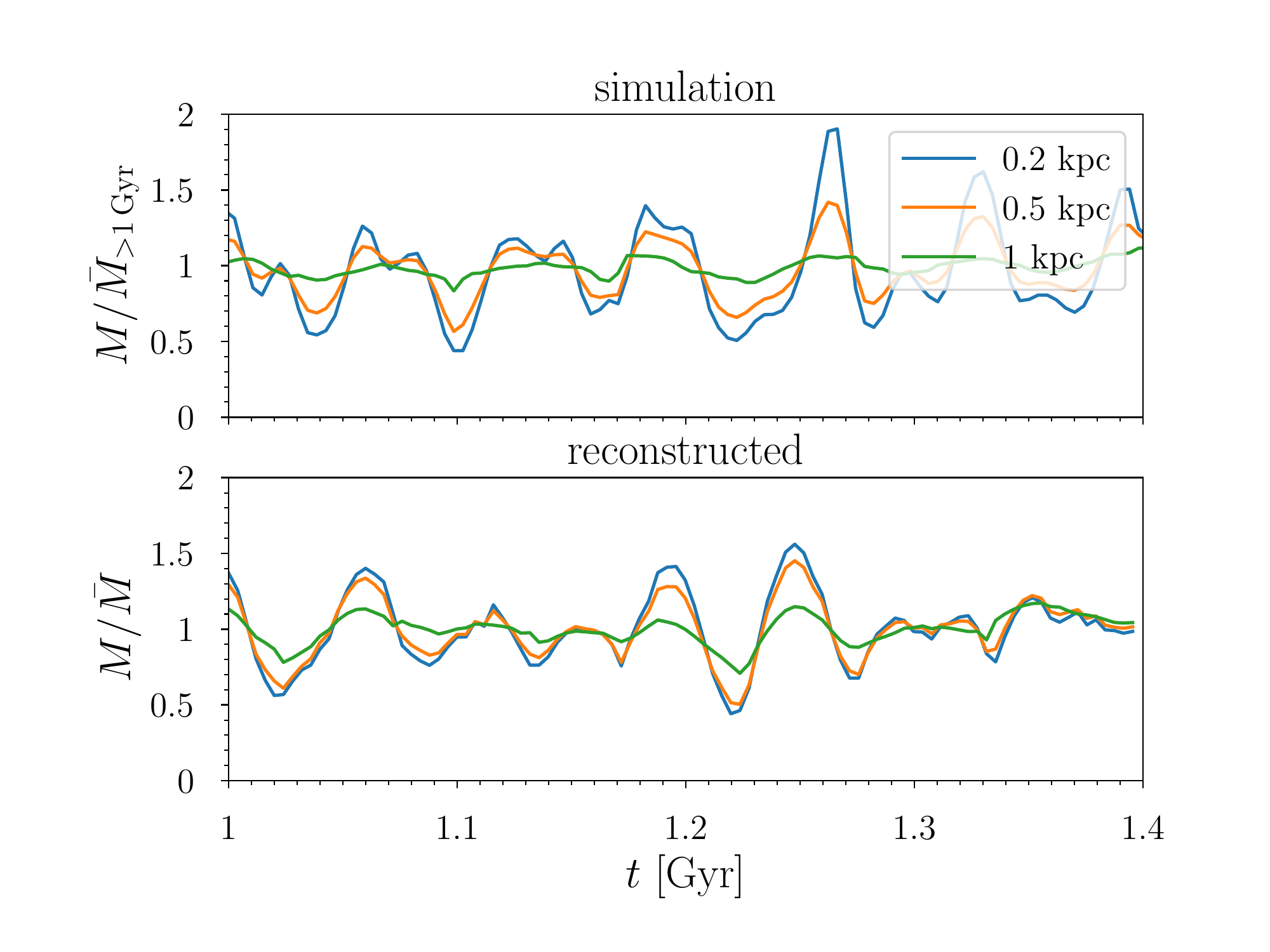}
\includegraphics[width=0.45\textwidth]{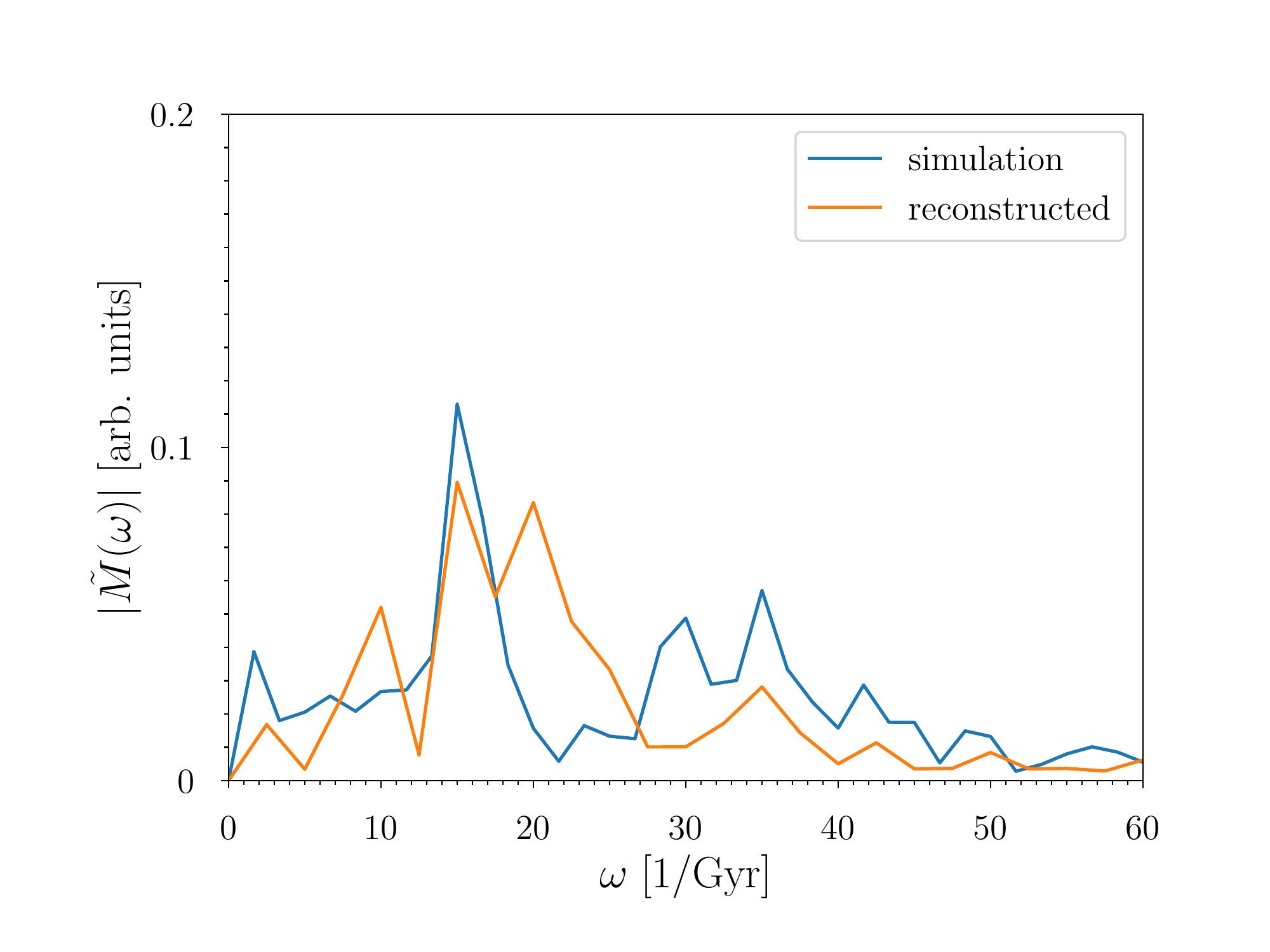} 
\caption{Comparison of soliton oscillations observed in the simulation and implied by the reconstructed wavefunction $\tilde{\psi}$.
Upper figure: the temporal variation of mass within several different radii from the soliton center.
Lower figure: Fourier coefficients $|\tilde{M}(\omega)|$ of the mass oscillation as a function of frequency $\omega$.
}
\label{fig:reconstruct_random}
\end{figure}

\begin{figure}[htpb]
\includegraphics[width=0.45\textwidth]{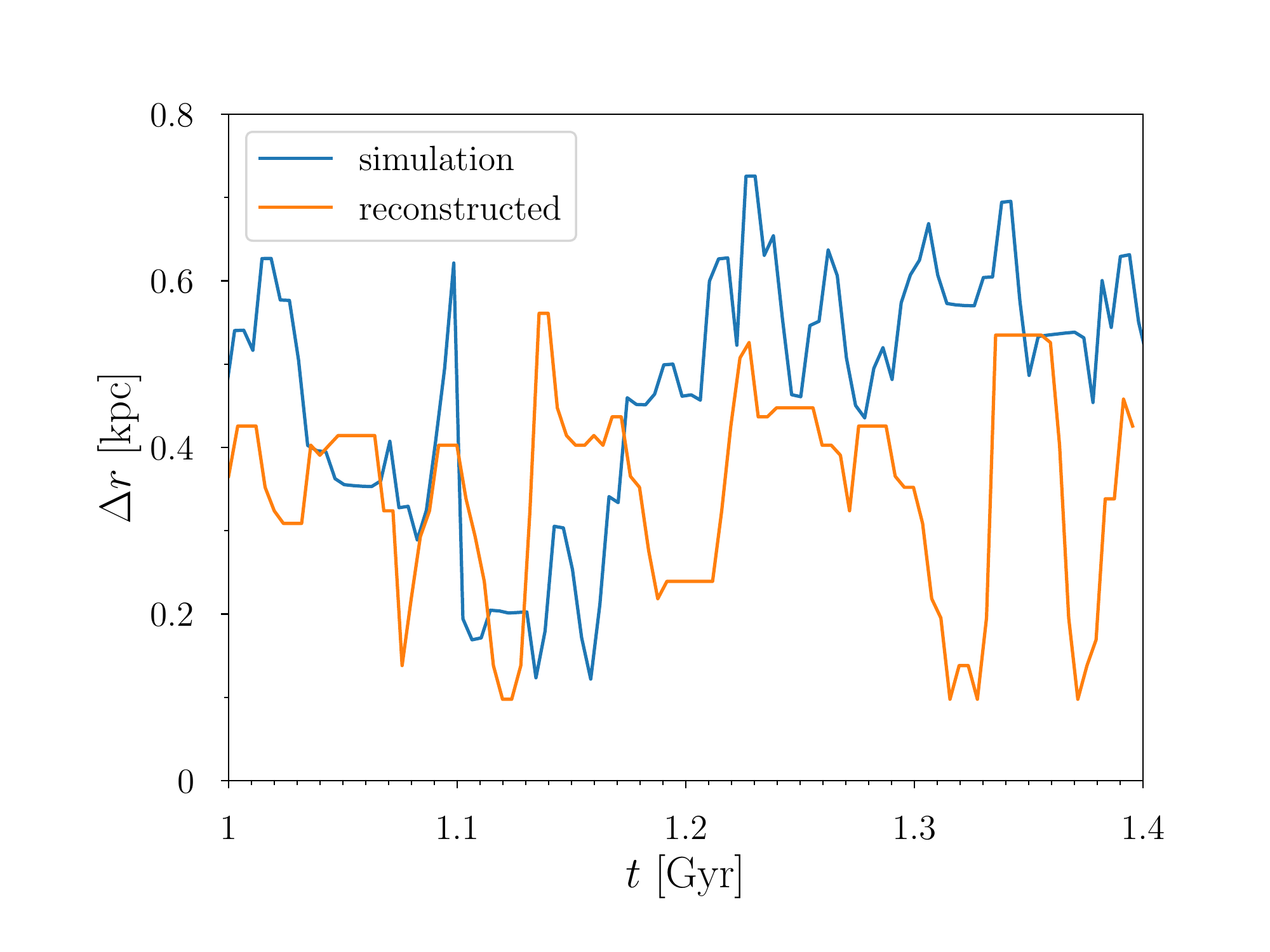} 
\caption{The distance $\Delta r$ of the soliton center from the halo center as a function of time (i.e. the soliton random walk), 
as measured from the soliton collision simulation and predicted by the reconstructed wavefunction.
}
\label{fig:soliton_pos}
\end{figure}

We repeat the same procedure as before, putting together $\tilde\psi$ (Equation (\ref{psidecomposeR})) using {\it fixed} coefficients $A_{nlm}$ (adopting their values at $1.2$ Gyr), including all modes with $n \le 3$. 
\footnote{In general, the energy eigenvalue depends on $n, l$ and $m$. Here, we compute the eigenmodes using the spherically averaged potential, and thus the energy depends only on $n$ and $l$.}
The top 2-panel plot in Figure \ref{fig:reconstruct_random} shows the temporal variations of the mass enclosed within a few different radii of the soliton, both measured from the simulation and implied by the reconstruction $\tilde\psi$. 
The bottom plot in Figure \ref{fig:reconstruct_random} shows the Fourier transform of the mass fluctuation curves. The peak frequency of $15.1$ Gyr$^{-1}$ can be identified with $\Delta E/\hbar$, with $\Delta E$ being the energy difference between the eigenstates $(200)$ and $(100)$. 

The reconstructed $\tilde\psi$ can also be used to predict how the peak of the density (where the soliton is) moves with respect to the halo center of mass. This is shown in 
Figure~\ref{fig:soliton_pos}, which shows the distance between the soliton center and the halo center as a function of time, from both the simulation and the reconstruction. 
As can be seen, the soliton random walks by an amount of the order of the soliton radius. 

For both the soliton oscillation and random walk phenomena, the reconstructed $\tilde\psi$, with its time dependent phase factor for each eigenmode, is sufficient to account for the broad features. 
In particular, the dominant oscillation frequency matches the energy difference of the two dominant $\ell=0$ modes. 
The $\ell \ne 0$ eigenmodes are suppressed at small radii (see Figure \ref{fig:modes_tophat}), but they do contribute to fluctuations at some level. 
In particular, the soliton random walk excursions are driven by them. 

\section{Discussion and Conclusion}

We have described two numerical experiments to study the oscillations of the solitonic core in FDM halos.
One is from spherically symmetric initial data and the other is from the merger of several randomly placed seed solitons. 
In both cases, persistent order unity oscillations are observed close to the central regions, even after the overall halo appears to have virialized. 
In addition, for the asymmetric case, the solitonic lump (or density peak) random walks. 
These reproduce the findings of \cite{Veltmaat:2018dfz,Schive:2019rrw}. 

\begin{figure}[htpb]
\includegraphics[width=0.45\textwidth]{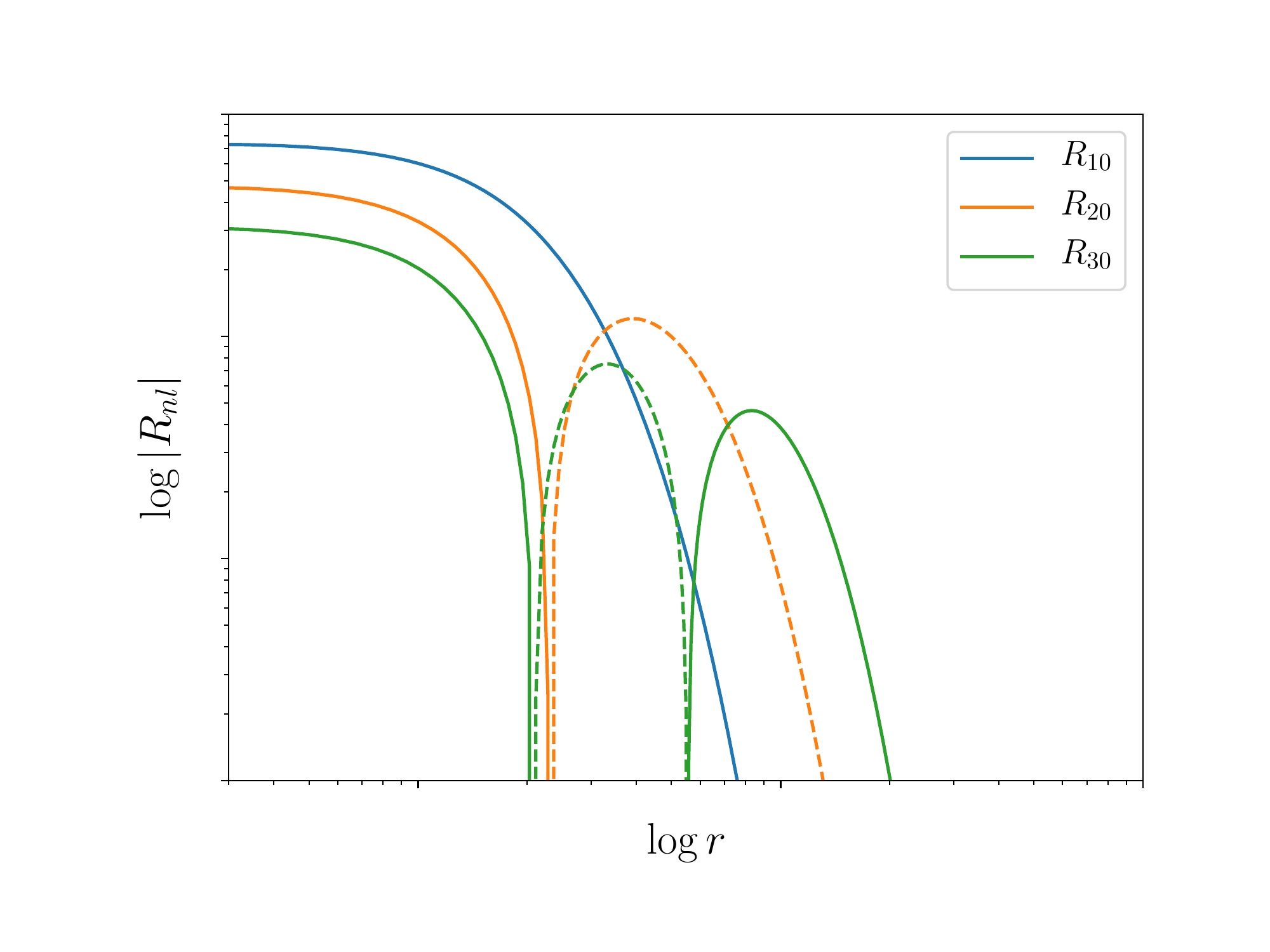} 
\caption{A schematic plot of the radial eigenfunction $R_{nl}$ for a few modes. 
Solid/dashed portions of the curves are for positive/negative values of $R_{nl}$. Not shown here are eigenstates with $l\ne 0$. 
The general trend holds regardless of $l$: excited states are more extended in radius compared to the ground state. 
The $l\ne0$ states are suppressed at small radii and hence not as important for soliton oscillations.
}
\label{illustration}
\end{figure}

Both can be understood as wave interference phenomena. 
As a first
approximation, one can model the halo wavefunction as in Equation
(\ref{psidecomposeR}). 
It bears repeating here:
\begin{equation}
\tilde \psi(r,\theta,\phi, t) = \sum_{n,l,m} A_{nlm}^{\rm fix} F_{nlm} (r,\theta,\phi) e^{-iE_{nl} t/\hbar} \, .
\end{equation}
Here, $F_{nlm}$'s are the energy eigenmodes of the gravitational
potential of the virialized halo. 
The potential does fluctuate in time (see Figures \ref{fig:profile_tophat} and \ref{fig:profile_random}), but its fluctuations are suppressed compared to those in density, for
the simple reason that gravity is a long range force, and the
potential is necessarily smoother than the density. 
Thus, precisely which potential we use to formulate the eigenmodes does not matter a great deal --- in practice, we use the potential averaged over a period of time. 
The coefficients $A_{nlm}^{\rm fix}$'s are the superposition amplitudes for the eigenmodes, and once again, the precise time at which they are fixed does not matter a great deal. 
Once this is done, the only time dependence of $\tilde\psi$ arises from the energy dependent phase factor for each mode.
The situation is schematically illustrated in Figure \ref{illustration}. 
When such a superposition of eigenmodes is ``squared'' to obtain the density, it is the cross-terms that contribute to time variability, with frequencies set by the difference in energies between different modes. 
This has some quantitative success in matching what is observed in the simulations (Figures \ref{fig:reconstruct_tophat} and \ref{fig:reconstruct_random}).

It is worth stressing that while the agreement is reassuring, it is also not surprising that the order of magnitude should work out. 
This is because the dynamical time associated with the soliton is roughly 
\begin{equation}
    \sqrt{\frac{r_c^3}{G M_c}} \sim \frac{\hbar^3}{m_a^3 G^2 M_c^2} \, ,
\end{equation}
(see relation between soliton mass $M_c$ and radius $r_c$ given in Section~\ref{sec:level1}). 
This is also roughly $\hbar/E$ where 
\begin{equation}
    E \sim \frac{m_a GM_c}{r_c} \sim \frac{m_a^3 G^2 M_c^2}{\hbar^2} \, ,
\end{equation} 
is the ground state soliton energy. 

The wave interference model presented here can be thought of as a
first approximation. 
There are quite a few interesting phenomena that can be investigated building on this approximation. 
They involve probing the temporal variability in the coefficients $A_{nlm}$'s, precisely what is ignored in the simplified model.
For instance, the phenomenon of soliton condensation has to do with the growth of $A_{100}$. 
If the initial $A_{100}$ were small relative to the other amplitudes, how does it grow with time? And what determines the ultimate ratio of this ground state amplitude relative to others? 
It is the relative contributions of the different states that determine not only the overall halo structure, but also the oscillation and random walk amplitudes. 
An investigation along this line, reminiscent of \cite{Levkov:2018kau}, would be interesting.

Another interesting application is the study of tidal disruption.
Imagine the halo of interest is in some tidal environment, much like a satellite galaxy such as Eridanus II residing in the Milky Way. Let's represent this satellite halo as a superposition of
eigenstates much as in Figure \ref{illustration}. 
Tidal stripping preferentially removes the outer parts of the halo
where the density is lower. 
Keep in mind that it is the excited states that contribute most to the outer portions of the halo.
\footnote{Classic tidal disruption is described by a tidal radius within which the satellite halo is protected from disruption. The wave nature of FDM modifies it, in that no region of the satellite halo is completely safe from tidal stripping \cite{Hui:2016ltb}. 
Nonetheless, tidal disruption remains more efficient in the outer parts of the halo where the density is lower.}
Meanwhile, as the overall gravitational potential of the satellite halo evolves due to mass loss, the ground state evolves adiabatically in response.
The net effect should be a reduction of the amplitudes of the excited states relative to the ground state. 
Thus, we expect diminished soliton oscillations and random walk excursions.
This appears to be what is seen in simulations by Schive, Chiueh, and Broadhurst \cite{Schive:2019rrw}, who also considered the implications for deriving FDM constraints from stellar heating.
This whole discussion ignores the fact that the waves associated with the parent halo also superimpose with those of the satellite. 
It is likely that the parent waves have a relative small impact on the satellite, but it would be useful to verify this with further
simulations.

\acknowledgements
We thank Neal Dalal and Hsi-Yu Schive for useful discussions.
X.L is supported by the Natural Sciences and Engineering Research Council of Canada (NSERC), funding reference \#CITA 490888-16 and acknowledges the support of computational resources provided by Compute Ontario and Compute Canada.
L.H. is supported by a Simons Fellowship and the Department of Energy DE-SC0011941.
T.D.Y. is supported through the NSF Graduate Research Fellowship (DGE-1644869). 
Research at Perimeter Institute is supported in part by the Government of Canada through the Department of Innovation, Science and Economic Development Canada and by the Province of Ontario through the Ministry of Colleges and Universities.

\bibliography{ms}

\begin{thebibliography}{32}%
\makeatletter
\providecommand \@ifxundefined [1]{%
 \@ifx{#1\undefined}
}%
\providecommand \@ifnum [1]{%
 \ifnum #1\expandafter \@firstoftwo
 \else \expandafter \@secondoftwo
 \fi
}%
\providecommand \@ifx [1]{%
 \ifx #1\expandafter \@firstoftwo
 \else \expandafter \@secondoftwo
 \fi
}%
\providecommand \natexlab [1]{#1}%
\providecommand \enquote  [1]{``#1''}%
\providecommand \bibnamefont  [1]{#1}%
\providecommand \bibfnamefont [1]{#1}%
\providecommand \citenamefont [1]{#1}%
\providecommand \href@noop [0]{\@secondoftwo}%
\providecommand \href [0]{\begingroup \@sanitize@url \@href}%
\providecommand \@href[1]{\@@startlink{#1}\@@href}%
\providecommand \@@href[1]{\endgroup#1\@@endlink}%
\providecommand \@sanitize@url [0]{\catcode `\\12\catcode `\$12\catcode
  `\&12\catcode `\#12\catcode `\^12\catcode `\_12\catcode `\%12\relax}%
\providecommand \@@startlink[1]{}%
\providecommand \@@endlink[0]{}%
\providecommand \url  [0]{\begingroup\@sanitize@url \@url }%
\providecommand \@url [1]{\endgroup\@href {#1}{\urlprefix }}%
\providecommand \urlprefix  [0]{URL }%
\providecommand \Eprint [0]{\href }%
\providecommand \doibase [0]{http://dx.doi.org/}%
\providecommand \selectlanguage [0]{\@gobble}%
\providecommand \bibinfo  [0]{\@secondoftwo}%
\providecommand \bibfield  [0]{\@secondoftwo}%
\providecommand \translation [1]{[#1]}%
\providecommand \BibitemOpen [0]{}%
\providecommand \bibitemStop [0]{}%
\providecommand \bibitemNoStop [0]{.\EOS\space}%
\providecommand \EOS [0]{\spacefactor3000\relax}%
\providecommand \BibitemShut  [1]{\csname bibitem#1\endcsname}%
\let\auto@bib@innerbib\@empty
\bibitem [{\citenamefont {Hu}\ \emph {et~al.}(2000)\citenamefont {Hu},
  \citenamefont {Barkana},\ and\ \citenamefont {Gruzinov}}]{Hu:2000ke}%
  \BibitemOpen
  \bibfield  {author} {\bibinfo {author} {\bibfnamefont {W.}~\bibnamefont
  {Hu}}, \bibinfo {author} {\bibfnamefont {R.}~\bibnamefont {Barkana}}, \ and\
  \bibinfo {author} {\bibfnamefont {A.}~\bibnamefont {Gruzinov}},\ }\href
  {\doibase 10.1103/PhysRevLett.85.1158} {\bibfield  {journal} {\bibinfo
  {journal} {Phys. Rev. Lett.}\ }\textbf {\bibinfo {volume} {85}},\ \bibinfo
  {pages} {1158} (\bibinfo {year} {2000})},\ \Eprint
  {http://arxiv.org/abs/astro-ph/0003365} {arXiv:astro-ph/0003365 [astro-ph]}
  \BibitemShut {NoStop}%
\bibitem [{\citenamefont {Baldeschi}\ \emph {et~al.}(1983)\citenamefont
  {Baldeschi}, \citenamefont {Ruffini},\ and\ \citenamefont
  {Gelmini}}]{Baldeschi:1983mq}%
  \BibitemOpen
  \bibfield  {author} {\bibinfo {author} {\bibfnamefont {M.~R.}\ \bibnamefont
  {Baldeschi}}, \bibinfo {author} {\bibfnamefont {R.}~\bibnamefont {Ruffini}},
  \ and\ \bibinfo {author} {\bibfnamefont {G.~B.}\ \bibnamefont {Gelmini}},\
  }\href {\doibase 10.1016/0370-2693(83)90688-3} {\bibfield  {journal}
  {\bibinfo  {journal} {Phys. Lett.}\ }\textbf {\bibinfo {volume} {122B}},\
  \bibinfo {pages} {221} (\bibinfo {year} {1983})}\BibitemShut {NoStop}%
\bibitem [{\citenamefont {Turner}(1983)}]{Turner:1983he}%
  \BibitemOpen
  \bibfield  {author} {\bibinfo {author} {\bibfnamefont {M.~S.}\ \bibnamefont
  {Turner}},\ }\href {\doibase 10.1103/PhysRevD.28.1243} {\bibfield  {journal}
  {\bibinfo  {journal} {Phys. Rev.}\ }\textbf {\bibinfo {volume} {D28}},\
  \bibinfo {pages} {1243} (\bibinfo {year} {1983})}\BibitemShut {NoStop}%
\bibitem [{\citenamefont {Press}\ \emph {et~al.}(1990)\citenamefont {Press},
  \citenamefont {Ryden},\ and\ \citenamefont {Spergel}}]{Press:1989id}%
  \BibitemOpen
  \bibfield  {author} {\bibinfo {author} {\bibfnamefont {W.~H.}\ \bibnamefont
  {Press}}, \bibinfo {author} {\bibfnamefont {B.~S.}\ \bibnamefont {Ryden}}, \
  and\ \bibinfo {author} {\bibfnamefont {D.~N.}\ \bibnamefont {Spergel}},\
  }\href {\doibase 10.1103/PhysRevLett.64.1084} {\bibfield  {journal} {\bibinfo
   {journal} {Phys. Rev. Lett.}\ }\textbf {\bibinfo {volume} {64}},\ \bibinfo
  {pages} {1084} (\bibinfo {year} {1990})}\BibitemShut {NoStop}%
\bibitem [{\citenamefont {Sin}(1994)}]{Sin:1992bg}%
  \BibitemOpen
  \bibfield  {author} {\bibinfo {author} {\bibfnamefont {S.-J.}\ \bibnamefont
  {Sin}},\ }\href {\doibase 10.1103/PhysRevD.50.3650} {\bibfield  {journal}
  {\bibinfo  {journal} {Phys. Rev.}\ }\textbf {\bibinfo {volume} {D50}},\
  \bibinfo {pages} {3650} (\bibinfo {year} {1994})},\ \Eprint
  {http://arxiv.org/abs/hep-ph/9205208} {arXiv:hep-ph/9205208 [hep-ph]}
  \BibitemShut {NoStop}%
\bibitem [{\citenamefont {Goodman}(2000)}]{Goodman:2000tg}%
  \BibitemOpen
  \bibfield  {author} {\bibinfo {author} {\bibfnamefont {J.}~\bibnamefont
  {Goodman}},\ }\href {\doibase 10.1016/S1384-1076(00)00015-4} {\bibfield
  {journal} {\bibinfo  {journal} {New Astron.}\ }\textbf {\bibinfo {volume}
  {5}},\ \bibinfo {pages} {103} (\bibinfo {year} {2000})},\ \Eprint
  {http://arxiv.org/abs/astro-ph/0003018} {arXiv:astro-ph/0003018 [astro-ph]}
  \BibitemShut {NoStop}%
\bibitem [{\citenamefont {Peebles}(2000)}]{Peebles:2000yy}%
  \BibitemOpen
  \bibfield  {author} {\bibinfo {author} {\bibfnamefont {P.~J.~E.}\
  \bibnamefont {Peebles}},\ }\href {\doibase 10.1086/312677} {\bibfield
  {journal} {\bibinfo  {journal} {Astrophys. J.}\ }\textbf {\bibinfo {volume}
  {534}},\ \bibinfo {pages} {L127} (\bibinfo {year} {2000})},\ \Eprint
  {http://arxiv.org/abs/astro-ph/0002495} {arXiv:astro-ph/0002495 [astro-ph]}
  \BibitemShut {NoStop}%
\bibitem [{\citenamefont {Lesgourgues}\ \emph {et~al.}(2002)\citenamefont
  {Lesgourgues}, \citenamefont {Arbey},\ and\ \citenamefont
  {Salati}}]{Lesgourgues:2002hk}%
  \BibitemOpen
  \bibfield  {author} {\bibinfo {author} {\bibfnamefont {J.}~\bibnamefont
  {Lesgourgues}}, \bibinfo {author} {\bibfnamefont {A.}~\bibnamefont {Arbey}},
  \ and\ \bibinfo {author} {\bibfnamefont {P.}~\bibnamefont {Salati}},\ }\href
  {\doibase 10.1016/S1387-6473(02)00247-6} {\bibfield  {journal} {\bibinfo
  {journal} {New Astron. Rev.}\ }\textbf {\bibinfo {volume} {46}},\ \bibinfo
  {pages} {791} (\bibinfo {year} {2002})}\BibitemShut {NoStop}%
\bibitem [{\citenamefont {Amendola}\ and\ \citenamefont
  {Barbieri}(2006)}]{Amendola:2005ad}%
  \BibitemOpen
  \bibfield  {author} {\bibinfo {author} {\bibfnamefont {L.}~\bibnamefont
  {Amendola}}\ and\ \bibinfo {author} {\bibfnamefont {R.}~\bibnamefont
  {Barbieri}},\ }\href {\doibase 10.1016/j.physletb.2006.08.069} {\bibfield
  {journal} {\bibinfo  {journal} {Phys. Lett.}\ }\textbf {\bibinfo {volume}
  {B642}},\ \bibinfo {pages} {192} (\bibinfo {year} {2006})},\ \Eprint
  {http://arxiv.org/abs/hep-ph/0509257} {arXiv:hep-ph/0509257 [hep-ph]}
  \BibitemShut {NoStop}%
\bibitem [{\citenamefont {Chavanis}(2011)}]{Chavanis:2011zi}%
  \BibitemOpen
  \bibfield  {author} {\bibinfo {author} {\bibfnamefont {P.-H.}\ \bibnamefont
  {Chavanis}},\ }\href {\doibase 10.1103/PhysRevD.84.043531} {\bibfield
  {journal} {\bibinfo  {journal} {Phys. Rev.}\ }\textbf {\bibinfo {volume}
  {D84}},\ \bibinfo {pages} {043531} (\bibinfo {year} {2011})},\ \Eprint
  {http://arxiv.org/abs/1103.2050} {arXiv:1103.2050 [astro-ph.CO]} \BibitemShut
  {NoStop}%
\bibitem [{\citenamefont {Arvanitaki}\ \emph {et~al.}(2010)\citenamefont
  {Arvanitaki}, \citenamefont {Dimopoulos}, \citenamefont {Dubovsky},
  \citenamefont {Kaloper},\ and\ \citenamefont
  {March-Russell}}]{Arvanitaki:2009fg}%
  \BibitemOpen
  \bibfield  {author} {\bibinfo {author} {\bibfnamefont {A.}~\bibnamefont
  {Arvanitaki}}, \bibinfo {author} {\bibfnamefont {S.}~\bibnamefont
  {Dimopoulos}}, \bibinfo {author} {\bibfnamefont {S.}~\bibnamefont
  {Dubovsky}}, \bibinfo {author} {\bibfnamefont {N.}~\bibnamefont {Kaloper}}, \
  and\ \bibinfo {author} {\bibfnamefont {J.}~\bibnamefont {March-Russell}},\
  }\href {\doibase 10.1103/PhysRevD.81.123530} {\bibfield  {journal} {\bibinfo
  {journal} {Phys. Rev.}\ }\textbf {\bibinfo {volume} {D81}},\ \bibinfo {pages}
  {123530} (\bibinfo {year} {2010})},\ \Eprint {http://arxiv.org/abs/0905.4720}
  {arXiv:0905.4720 [hep-th]} \BibitemShut {NoStop}%
\bibitem [{\citenamefont {Marsh}(2016)}]{Marsh:2015xka}%
  \BibitemOpen
  \bibfield  {author} {\bibinfo {author} {\bibfnamefont {D.~J.~E.}\
  \bibnamefont {Marsh}},\ }\href {\doibase 10.1016/j.physrep.2016.06.005}
  {\bibfield  {journal} {\bibinfo  {journal} {Phys. Rept.}\ }\textbf {\bibinfo
  {volume} {643}},\ \bibinfo {pages} {1} (\bibinfo {year} {2016})},\ \Eprint
  {http://arxiv.org/abs/1510.07633} {arXiv:1510.07633 [astro-ph.CO]}
  \BibitemShut {NoStop}%
\bibitem [{\citenamefont {Hui}\ \emph {et~al.}(2017)\citenamefont {Hui},
  \citenamefont {Ostriker}, \citenamefont {Tremaine},\ and\ \citenamefont
  {Witten}}]{Hui:2016ltb}%
  \BibitemOpen
  \bibfield  {author} {\bibinfo {author} {\bibfnamefont {L.}~\bibnamefont
  {Hui}}, \bibinfo {author} {\bibfnamefont {J.~P.}\ \bibnamefont {Ostriker}},
  \bibinfo {author} {\bibfnamefont {S.}~\bibnamefont {Tremaine}}, \ and\
  \bibinfo {author} {\bibfnamefont {E.}~\bibnamefont {Witten}},\ }\href
  {\doibase 10.1103/PhysRevD.95.043541} {\bibfield  {journal} {\bibinfo
  {journal} {Phys. Rev.}\ }\textbf {\bibinfo {volume} {D95}},\ \bibinfo {pages}
  {043541} (\bibinfo {year} {2017})},\ \Eprint
  {http://arxiv.org/abs/1610.08297} {arXiv:1610.08297 [astro-ph.CO]}
  \BibitemShut {NoStop}%
\bibitem [{\citenamefont {Madelung}(1927)}]{Madelung1927}%
  \BibitemOpen
  \bibfield  {author} {\bibinfo {author} {\bibfnamefont {E.}~\bibnamefont
  {Madelung}},\ }\href {\doibase 10.1007/BF01400372} {\bibfield  {journal}
  {\bibinfo  {journal} {Zeitschrift f{\"u}r Physik}\ }\textbf {\bibinfo
  {volume} {40}},\ \bibinfo {pages} {322} (\bibinfo {year} {1927})}\BibitemShut
  {NoStop}%
\bibitem [{\citenamefont {Hui}\ \emph {et~al.}(2020)\citenamefont {Hui},
  \citenamefont {Joyce}, \citenamefont {Landry},\ and\ \citenamefont
  {Li}}]{Hui:2020hbq}%
  \BibitemOpen
  \bibfield  {author} {\bibinfo {author} {\bibfnamefont {L.}~\bibnamefont
  {Hui}}, \bibinfo {author} {\bibfnamefont {A.}~\bibnamefont {Joyce}}, \bibinfo
  {author} {\bibfnamefont {M.~J.}\ \bibnamefont {Landry}}, \ and\ \bibinfo
  {author} {\bibfnamefont {X.}~\bibnamefont {Li}},\ }\href@noop {} {\
  (\bibinfo {year} {2020})},\ \Eprint {http://arxiv.org/abs/2004.01188}
  {arXiv:2004.01188 [astro-ph.CO]} \BibitemShut {NoStop}%
\bibitem [{\citenamefont {Schive}\ \emph
  {et~al.}(2014{\natexlab{a}})\citenamefont {Schive}, \citenamefont {Chiueh},\
  and\ \citenamefont {Broadhurst}}]{Schive:2014dra}%
  \BibitemOpen
  \bibfield  {author} {\bibinfo {author} {\bibfnamefont {H.-Y.}\ \bibnamefont
  {Schive}}, \bibinfo {author} {\bibfnamefont {T.}~\bibnamefont {Chiueh}}, \
  and\ \bibinfo {author} {\bibfnamefont {T.}~\bibnamefont {Broadhurst}},\
  }\href {\doibase 10.1038/nphys2996} {\bibfield  {journal} {\bibinfo
  {journal} {Nature Phys.}\ }\textbf {\bibinfo {volume} {10}},\ \bibinfo
  {pages} {496} (\bibinfo {year} {2014}{\natexlab{a}})},\ \Eprint
  {http://arxiv.org/abs/1406.6586} {arXiv:1406.6586 [astro-ph.GA]} \BibitemShut
  {NoStop}%
\bibitem [{\citenamefont {Schive}\ \emph
  {et~al.}(2014{\natexlab{b}})\citenamefont {Schive}, \citenamefont {Liao},
  \citenamefont {Woo}, \citenamefont {Wong}, \citenamefont {Chiueh},
  \citenamefont {Broadhurst},\ and\ \citenamefont {Hwang}}]{Schive:2014hza}%
  \BibitemOpen
  \bibfield  {author} {\bibinfo {author} {\bibfnamefont {H.-Y.}\ \bibnamefont
  {Schive}}, \bibinfo {author} {\bibfnamefont {M.-H.}\ \bibnamefont {Liao}},
  \bibinfo {author} {\bibfnamefont {T.-P.}\ \bibnamefont {Woo}}, \bibinfo
  {author} {\bibfnamefont {S.-K.}\ \bibnamefont {Wong}}, \bibinfo {author}
  {\bibfnamefont {T.}~\bibnamefont {Chiueh}}, \bibinfo {author} {\bibfnamefont
  {T.}~\bibnamefont {Broadhurst}}, \ and\ \bibinfo {author} {\bibfnamefont
  {W.~Y.~P.}\ \bibnamefont {Hwang}},\ }\href {\doibase
  10.1103/PhysRevLett.113.261302} {\bibfield  {journal} {\bibinfo  {journal}
  {Phys. Rev. Lett.}\ }\textbf {\bibinfo {volume} {113}},\ \bibinfo {pages}
  {261302} (\bibinfo {year} {2014}{\natexlab{b}})},\ \Eprint
  {http://arxiv.org/abs/1407.7762} {arXiv:1407.7762 [astro-ph.GA]} \BibitemShut
  {NoStop}%
\bibitem [{\citenamefont {Mocz}\ and\ \citenamefont
  {Succi}(2015)}]{Mocz:2015sda}%
  \BibitemOpen
  \bibfield  {author} {\bibinfo {author} {\bibfnamefont {P.}~\bibnamefont
  {Mocz}}\ and\ \bibinfo {author} {\bibfnamefont {S.}~\bibnamefont {Succi}},\
  }\href {\doibase 10.1103/PhysRevE.91.053304} {\bibfield  {journal} {\bibinfo
  {journal} {Phys. Rev.}\ }\textbf {\bibinfo {volume} {E91}},\ \bibinfo {pages}
  {053304} (\bibinfo {year} {2015})},\ \Eprint
  {http://arxiv.org/abs/1503.03869} {arXiv:1503.03869 [physics.comp-ph]}
  \BibitemShut {NoStop}%
\bibitem [{\citenamefont {Mocz}\ \emph {et~al.}(2017)\citenamefont {Mocz},
  \citenamefont {Vogelsberger}, \citenamefont {Robles}, \citenamefont {Zavala},
  \citenamefont {Boylan-Kolchin},\ and\ \citenamefont
  {Hernquist}}]{Mocz:2017wlg}%
  \BibitemOpen
  \bibfield  {author} {\bibinfo {author} {\bibfnamefont {P.}~\bibnamefont
  {Mocz}}, \bibinfo {author} {\bibfnamefont {M.}~\bibnamefont {Vogelsberger}},
  \bibinfo {author} {\bibfnamefont {V.}~\bibnamefont {Robles}}, \bibinfo
  {author} {\bibfnamefont {J.}~\bibnamefont {Zavala}}, \bibinfo {author}
  {\bibfnamefont {M.}~\bibnamefont {Boylan-Kolchin}}, \ and\ \bibinfo {author}
  {\bibfnamefont {L.}~\bibnamefont {Hernquist}},\ }\href {\doibase
  10.1093/mnras/stx1887} {\bibfield  {journal} {\bibinfo  {journal} {Mon. Not.
  Roy. Astron. Soc.}\ }\textbf {\bibinfo {volume} {471}},\ \bibinfo {pages}
  {4559} (\bibinfo {year} {2017})},\ \Eprint {http://arxiv.org/abs/1705.05845}
  {arXiv:1705.05845 [astro-ph.CO]} \BibitemShut {NoStop}%
\bibitem [{\citenamefont {Zhang}\ \emph {et~al.}(2016)\citenamefont {Zhang},
  \citenamefont {Tsai}, \citenamefont {Cheung},\ and\ \citenamefont
  {Chu}}]{Zhang:2016uiy}%
  \BibitemOpen
  \bibfield  {author} {\bibinfo {author} {\bibfnamefont {J.}~\bibnamefont
  {Zhang}}, \bibinfo {author} {\bibfnamefont {Y.-L.~S.}\ \bibnamefont {Tsai}},
  \bibinfo {author} {\bibfnamefont {K.}~\bibnamefont {Cheung}}, \ and\ \bibinfo
  {author} {\bibfnamefont {M.-C.}\ \bibnamefont {Chu}},\ }\href@noop {} {\
  (\bibinfo {year} {2016})},\ \Eprint {http://arxiv.org/abs/1611.00892}
  {arXiv:1611.00892 [astro-ph.CO]} \BibitemShut {NoStop}%
\bibitem [{\citenamefont {Schwabe}\ \emph {et~al.}(2016)\citenamefont
  {Schwabe}, \citenamefont {Niemeyer},\ and\ \citenamefont
  {Engels}}]{Schwabe:2016rze}%
  \BibitemOpen
  \bibfield  {author} {\bibinfo {author} {\bibfnamefont {B.}~\bibnamefont
  {Schwabe}}, \bibinfo {author} {\bibfnamefont {J.~C.}\ \bibnamefont
  {Niemeyer}}, \ and\ \bibinfo {author} {\bibfnamefont {J.~F.}\ \bibnamefont
  {Engels}},\ }\href {\doibase 10.1103/PhysRevD.94.043513} {\bibfield
  {journal} {\bibinfo  {journal} {Phys. Rev.}\ }\textbf {\bibinfo {volume}
  {D94}},\ \bibinfo {pages} {043513} (\bibinfo {year} {2016})},\ \Eprint
  {http://arxiv.org/abs/1606.05151} {arXiv:1606.05151 [astro-ph.CO]}
  \BibitemShut {NoStop}%
\bibitem [{\citenamefont {Li}\ \emph {et~al.}(2019)\citenamefont {Li},
  \citenamefont {Hui},\ and\ \citenamefont {Bryan}}]{Li:2018kyk}%
  \BibitemOpen
  \bibfield  {author} {\bibinfo {author} {\bibfnamefont {X.}~\bibnamefont
  {Li}}, \bibinfo {author} {\bibfnamefont {L.}~\bibnamefont {Hui}}, \ and\
  \bibinfo {author} {\bibfnamefont {G.~L.}\ \bibnamefont {Bryan}},\ }\href
  {\doibase 10.1103/PhysRevD.99.063509} {\bibfield  {journal} {\bibinfo
  {journal} {Phys. Rev. D}\ }\textbf {\bibinfo {volume} {99}},\ \bibinfo
  {pages} {063509} (\bibinfo {year} {2019})},\ \Eprint
  {http://arxiv.org/abs/1810.01915} {arXiv:1810.01915 [astro-ph.CO]}
  \BibitemShut {NoStop}%
\bibitem [{\citenamefont {Nori}\ and\ \citenamefont
  {Baldi}(2018)}]{Nori:2018hud}%
  \BibitemOpen
  \bibfield  {author} {\bibinfo {author} {\bibfnamefont {M.}~\bibnamefont
  {Nori}}\ and\ \bibinfo {author} {\bibfnamefont {M.}~\bibnamefont {Baldi}},\
  }\href {\doibase 10.1093/mnras/sty1224} {\bibfield  {journal} {\bibinfo
  {journal} {Mon. Not. Roy. Astron. Soc.}\ }\textbf {\bibinfo {volume} {478}},\
  \bibinfo {pages} {3935} (\bibinfo {year} {2018})},\ \Eprint
  {http://arxiv.org/abs/1801.08144} {arXiv:1801.08144 [astro-ph.CO]}
  \BibitemShut {NoStop}%
\bibitem [{\citenamefont {Veltmaat}\ and\ \citenamefont
  {Niemeyer}(2016)}]{Veltmaat:2016rxo}%
  \BibitemOpen
  \bibfield  {author} {\bibinfo {author} {\bibfnamefont {J.}~\bibnamefont
  {Veltmaat}}\ and\ \bibinfo {author} {\bibfnamefont {J.~C.}\ \bibnamefont
  {Niemeyer}},\ }\href {\doibase 10.1103/PhysRevD.94.123523} {\bibfield
  {journal} {\bibinfo  {journal} {Phys. Rev.}\ }\textbf {\bibinfo {volume}
  {D94}},\ \bibinfo {pages} {123523} (\bibinfo {year} {2016})},\ \Eprint
  {http://arxiv.org/abs/1608.00802} {arXiv:1608.00802 [astro-ph.CO]}
  \BibitemShut {NoStop}%
\bibitem [{\citenamefont {Veltmaat}\ \emph {et~al.}(2018)\citenamefont
  {Veltmaat}, \citenamefont {Niemeyer},\ and\ \citenamefont
  {Schwabe}}]{Veltmaat:2018dfz}%
  \BibitemOpen
  \bibfield  {author} {\bibinfo {author} {\bibfnamefont {J.}~\bibnamefont
  {Veltmaat}}, \bibinfo {author} {\bibfnamefont {J.~C.}\ \bibnamefont
  {Niemeyer}}, \ and\ \bibinfo {author} {\bibfnamefont {B.}~\bibnamefont
  {Schwabe}},\ }\href {\doibase 10.1103/PhysRevD.98.043509} {\bibfield
  {journal} {\bibinfo  {journal} {Phys. Rev.}\ }\textbf {\bibinfo {volume}
  {D98}},\ \bibinfo {pages} {043509} (\bibinfo {year} {2018})},\ \Eprint
  {http://arxiv.org/abs/1804.09647} {arXiv:1804.09647 [astro-ph.CO]}
  \BibitemShut {NoStop}%
\bibitem [{\citenamefont {Schwabe}\ \emph {et~al.}(2020)\citenamefont
  {Schwabe}, \citenamefont {Gosenca}, \citenamefont {Behrens}, \citenamefont
  {Niemeyer},\ and\ \citenamefont {Easther}}]{Schwabe:2020eac}%
  \BibitemOpen
  \bibfield  {author} {\bibinfo {author} {\bibfnamefont {B.}~\bibnamefont
  {Schwabe}}, \bibinfo {author} {\bibfnamefont {M.}~\bibnamefont {Gosenca}},
  \bibinfo {author} {\bibfnamefont {C.}~\bibnamefont {Behrens}}, \bibinfo
  {author} {\bibfnamefont {J.~C.}\ \bibnamefont {Niemeyer}}, \ and\ \bibinfo
  {author} {\bibfnamefont {R.}~\bibnamefont {Easther}},\ }\href {\doibase
  10.1103/PhysRevD.102.083518} {\bibfield  {journal} {\bibinfo  {journal}
  {Phys. Rev. D}\ }\textbf {\bibinfo {volume} {102}},\ \bibinfo {pages}
  {083518} (\bibinfo {year} {2020})},\ \Eprint
  {http://arxiv.org/abs/2007.08256} {arXiv:2007.08256 [astro-ph.CO]}
  \BibitemShut {NoStop}%
\bibitem [{\citenamefont {Navarro}\ \emph {et~al.}(1997)\citenamefont
  {Navarro}, \citenamefont {Frenk},\ and\ \citenamefont
  {White}}]{Navarro:1996gj}%
  \BibitemOpen
  \bibfield  {author} {\bibinfo {author} {\bibfnamefont {J.~F.}\ \bibnamefont
  {Navarro}}, \bibinfo {author} {\bibfnamefont {C.~S.}\ \bibnamefont {Frenk}},
  \ and\ \bibinfo {author} {\bibfnamefont {S.~D.}\ \bibnamefont {White}},\
  }\href {\doibase 10.1086/304888} {\bibfield  {journal} {\bibinfo  {journal}
  {Astrophys. J.}\ }\textbf {\bibinfo {volume} {490}},\ \bibinfo {pages} {493}
  (\bibinfo {year} {1997})},\ \Eprint {http://arxiv.org/abs/astro-ph/9611107}
  {arXiv:astro-ph/9611107} \BibitemShut {NoStop}%
\bibitem [{\citenamefont {Marsh}\ and\ \citenamefont
  {Niemeyer}(2019)}]{Marsh:2018zyw}%
  \BibitemOpen
  \bibfield  {author} {\bibinfo {author} {\bibfnamefont {D.~J.}\ \bibnamefont
  {Marsh}}\ and\ \bibinfo {author} {\bibfnamefont {J.~C.}\ \bibnamefont
  {Niemeyer}},\ }\href {\doibase 10.1103/PhysRevLett.123.051103} {\bibfield
  {journal} {\bibinfo  {journal} {Phys. Rev. Lett.}\ }\textbf {\bibinfo
  {volume} {123}},\ \bibinfo {pages} {051103} (\bibinfo {year} {2019})},\
  \Eprint {http://arxiv.org/abs/1810.08543} {arXiv:1810.08543 [astro-ph.CO]}
  \BibitemShut {NoStop}%
\bibitem [{\citenamefont {Schive}\ \emph {et~al.}(2020)\citenamefont {Schive},
  \citenamefont {Chiueh},\ and\ \citenamefont {Broadhurst}}]{Schive:2019rrw}%
  \BibitemOpen
  \bibfield  {author} {\bibinfo {author} {\bibfnamefont {H.-Y.}\ \bibnamefont
  {Schive}}, \bibinfo {author} {\bibfnamefont {T.}~\bibnamefont {Chiueh}}, \
  and\ \bibinfo {author} {\bibfnamefont {T.}~\bibnamefont {Broadhurst}},\
  }\href {\doibase 10.1103/PhysRevLett.124.201301} {\bibfield  {journal}
  {\bibinfo  {journal} {Phys. Rev. Lett.}\ }\textbf {\bibinfo {volume} {124}},\
  \bibinfo {pages} {201301} (\bibinfo {year} {2020})},\ \Eprint
  {http://arxiv.org/abs/1912.09483} {arXiv:1912.09483 [astro-ph.GA]}
  \BibitemShut {NoStop}%
\bibitem [{\citenamefont {Lin}\ \emph {et~al.}(2018)\citenamefont {Lin},
  \citenamefont {Schive}, \citenamefont {Wong},\ and\ \citenamefont
  {Chiueh}}]{Lin:2018whl}%
  \BibitemOpen
  \bibfield  {author} {\bibinfo {author} {\bibfnamefont {S.-C.}\ \bibnamefont
  {Lin}}, \bibinfo {author} {\bibfnamefont {H.-Y.}\ \bibnamefont {Schive}},
  \bibinfo {author} {\bibfnamefont {S.-K.}\ \bibnamefont {Wong}}, \ and\
  \bibinfo {author} {\bibfnamefont {T.}~\bibnamefont {Chiueh}},\ }\href
  {\doibase 10.1103/PhysRevD.97.103523} {\bibfield  {journal} {\bibinfo
  {journal} {Phys. Rev.}\ }\textbf {\bibinfo {volume} {D97}},\ \bibinfo {pages}
  {103523} (\bibinfo {year} {2018})},\ \Eprint
  {http://arxiv.org/abs/1801.02320} {arXiv:1801.02320 [astro-ph.CO]}
  \BibitemShut {NoStop}%
\bibitem [{\citenamefont {Padmanabhan}\ and\ \citenamefont
  {et~al.}()}]{Padmanabhan:2020}%
  \BibitemOpen
  \bibfield  {author} {\bibinfo {author} {\bibfnamefont {N.}~\bibnamefont
  {Padmanabhan}}\ and\ \bibinfo {author} {\bibnamefont {et~al.}},\ }\href@noop
  {} {\bibinfo  {journal} {in preparation}\ }\BibitemShut {NoStop}%
\bibitem [{\citenamefont {Levkov}\ \emph {et~al.}(2018)\citenamefont {Levkov},
  \citenamefont {Panin},\ and\ \citenamefont {Tkachev}}]{Levkov:2018kau}%
  \BibitemOpen
\bibfield  {journal} {  }\bibfield  {author} {\bibinfo {author} {\bibfnamefont
  {D.}~\bibnamefont {Levkov}}, \bibinfo {author} {\bibfnamefont
  {A.}~\bibnamefont {Panin}}, \ and\ \bibinfo {author} {\bibfnamefont
  {I.}~\bibnamefont {Tkachev}},\ }\href {\doibase
  10.1103/PhysRevLett.121.151301} {\bibfield  {journal} {\bibinfo  {journal}
  {Phys. Rev. Lett.}\ }\textbf {\bibinfo {volume} {121}},\ \bibinfo {pages}
  {151301} (\bibinfo {year} {2018})},\ \Eprint
  {http://arxiv.org/abs/1804.05857} {arXiv:1804.05857 [astro-ph.CO]}
  \BibitemShut {NoStop}%
\end{thebibliography}%

\end{document}